\documentclass[12pt,letterpaper]{article}

\usepackage{thmtools}
\usepackage{thm-restate}
\usepackage{cleveref}

\usepackage{mathrsfs}           
\usepackage{vmargin,fancyhdr}   
\usepackage{algorithm}
\usepackage{algorithmic}

\usepackage{enumerate}

\usepackage{amsmath,amssymb}    
\usepackage{verbatim}           
\usepackage{xspace}             
\usepackage{graphicx,float}     
\usepackage{ifthen,calc}        
\usepackage{textcomp}           
\usepackage{fancybox}           
\usepackage{hhline}             
\usepackage{float}              

\usepackage[rflt]{floatflt}     
\usepackage[small,compact]{titlesec}
\usepackage{setspace}
\usepackage{subfigure}

\usepackage{color}
\definecolor{ForestGreen}{rgb}{0.1333,0.5451,0.1333}
\usepackage[letterpaper,
            colorlinks,linkcolor=ForestGreen,citecolor=ForestGreen,
            backref,
            bookmarks,bookmarksopen,bookmarksnumbered]
           {hyperref}
\usepackage{thumbpdf}

\setpapersize{USletter}
\setmarginsrb{.75in}{.5in}        
             {.75in}{.5in}        
             {.25in}{.25in}     
             {.25in}{.5in}      
\newcommand{\showccc}[0]{0}
\newcommand{\ccc}[2][nothing]{
  \ifthenelse{\showccc=0}{}{
    \ensuremath{^{\Lsh\Rsh}}\marginpar{\raggedright\tiny\textsf{%
        \ifthenelse{\equal{#1}{nothing}}{}{\textbf{#1}\\}#2}}}}

\newcounter{hours}\newcounter{minutes}
\newcommand{\hhmm}{%
  \setcounter{hours}{\time/60}%
  \setcounter{minutes}{\time-\value{hours}*60}%
  \ifthenelse{\value{hours}<10}{0}{}\thehours:%
  \ifthenelse{\value{minutes}<10}{0}{}\theminutes}
\ifthenelse{\showccc=0}{\rhead{}}{\rhead{\today \ [\hhmm]}}
\newtheorem{theorem}{Theorem}[section]

\newtheorem{corollary}[theorem]{Corollary}
\newtheorem{definition}[theorem]{Definition}

\newtheorem{lemma}[theorem]{Lemma}
\newtheorem{fact}[theorem]{Fact}

\newcommand{\Proof}[0]{\smallskip\noindent\textit{\textbf{Proof}}\quad}
\newcommand{\Proofof}[1]{\smallskip\noindent\textit{\textbf{Proof of #1:}}\quad}

\newcommand{\QED}[0]{\hfill\ensuremath{\blacksquare}\medspace}

\floatstyle{ruled}
\newfloat{algo}{tbp}{lop}
\floatname{algo}{Algorithm}

\usepackage{float}
\floatstyle{plain}\newfloat{myfig}{t}{figs}[section]
\floatname{myfig}{\textsc{Figure}}
\floatstyle{plain}\newfloat{myalg}{H}{algs}[section]
\floatname{myalg}{}
\title{Stretching Stretch\thanks{Work supported in part by NSF grants CCF-1018463 and CCF-1065106.}}

\author{
  Michael B. Cohen\\
  M.I.T.\thanks{Part of this work was done while at CMU}\\
  \texttt{micohen@mit.edu}
  \and
  Gary L. Miller\\
  Carnegie Mellon University\\
  \texttt{glmiller@cs.cmu.edu}
  \and
  Jakub W. Pachocki\\
  Carnegie Mellon University\\
  \texttt{pachocki@cs.cmu.edu}
  \and
  Richard Peng\\
  M.I.T.\thanks{Part of this work was done while at CMU and was supported by a Microsoft Research PhD Fellowship}\\
  \texttt{rpeng@mit.edu}
  \and
  Shen Chen Xu\\
  Carnegie Mellon University\\
  \texttt{shenchex@cs.cmu.edu}
}

\newcommand{\expct}{\ensuremath{\mathbb{E}}}
\newcommand{\disjunion}{\ensuremath{\mathaccent\cdot\cup}}

\newcommand{\defeq}{\buildrel \text{d{}ef}\over =}
\newcommand{\Oh}{\ensuremath{\mathcal{O}}}

\newcommand{\str}{\ensuremath{\textbf{STR}}}

\newcommand{\length}{\ensuremath{d}}
\newcommand{\weight}{\ensuremath{w}}
\newcommand{\laplacian}{\ensuremath{{L}}}

\newcommand{\proj}{\ensuremath{{\Pi}}}
\newcommand{\mata}{\ensuremath{{A}}}
\newcommand{\matb}{\ensuremath{{B}}}

\newcommand{\vecb}{\ensuremath{\textbf{b}}}
\newcommand{\vecx}{\ensuremath{\textbf{x}}}

\newcommand{\ones}{\vec{1}}

\newcommand{\poly}{\ensuremath{\textbf{poly}}}

\renewcommand{\AA}{\ensuremath{\textbf{A}}}
\newcommand{\BB}{\ensuremath{\textbf{B}}}
\newcommand{\dd}{\ensuremath{\textbf{d}}}

\renewcommand{\mit}[1]{\mathit{#1}}
\newcommand{\scope}{\ensuremath{\mit{scope}}}

\begin{document}

\maketitle

\begin{abstract}
We give a generalized definition of stretch that simplifies the efficient
construction of low-stretch embeddings suitable for graph algorithms.
The generalization, based on discounting highly stretched edges by
taking their $p\textsuperscript{th}$ power for some $0 < p < 1$,
is directly related to performances of existing algorithms.
This discounting of high-stretch edges allows us to treat
many classes of edges with coarser granularity.
It leads to a two-pass approach that combines bottom-up clustering
and top-down decompositions to construct these embeddings
in $\Oh(m\log\log{n})$ time.
Our algorithm parallelizes readily and can also produce
generalizations of low-stretch subgraphs.
\end{abstract}

\section{Introduction}
\label{sec:intro}

Over the last few years substantial progress has been made on a large
class of graph theoretic optimization problems.
We now know substantially better asymptotic running time bounds
and parallelizations for approximate undirected maximum flow/minimum cut~\cite{Madry10b,ChristianoKMST10,LeeRS13,KelnerLOS13,Sherman13}, 
bipartite matching~\cite{Madry13},
minimum cost maximum flow~\cite{DaitchS08}, 
minimum energy flows~\cite{SpielmanT04,KoutisMP11,KelnerOSZ13,CohenFMNPW14},
and graph partitioning~\cite{Sherman09,OrecchiaSV12}.  
One commonality of all these new algorithms is that they either
explicitly find low-stretch spanning trees
or call an algorithm that at least at present uses these trees.

The fastest  known algorithm for generating these trees, 
due to Abraham and Neiman runs in $\Oh(m\log n \log \log n)$ time~\cite{AbrahamN12}.
Among the problems listed above, this running time is only the bottleneck
for the minimum energy flow problem and its dual, solving
symmetric diagonally dominant linear systems.
However, there is optimism that all of the above problems can
be solved in $o(m\log n)$ time, in which case finding these trees
becomes a bottleneck as well.
The main question we address in this paper is finding algorithms for
constructing even better trees in $\Oh(m)$ time.
Unfortunately, this remains an open question.

This paper removes the tree construction obstacle from $o(m\log{n})$
time algorithms for solving SDD systems,
as well as other graph optimization problems.
We give two modifications to the definition of low stretch spanning trees
that can simplify and speed up their construction.
Firstly, we allow additional vertices in the tree, leading to a Steiner tree.
This avoids the need for the complex graph decomposition scheme of \cite{AbrahamN12}.
Secondly, we discount the cost of high-stretch edges in ways that more
accurately reflect how these trees are used.  This allows the algorithm to be more
``forgetful,'' and is crucial to our speedup.

Throughout this paper we let $G=(V,E,l)$ be a graph with edge lengths $l(e)$,
and $T = (V_T, E_T, l_T)$ to denote the trees that we consider.
In previous works on low stretch spanning trees, $T$ was required to
be a subgraph of $G$ in the weighted sense.
In other words, $E_T \subseteq E$, and $l_T(e) = l(e)$ for all  $e \in E_T$.
We relax this condition by only requiring edge lengths in $T$ to be not too
short with respect to $G$ through the notion of embeddability,
which we formalize in Section~\ref{sec:background}.

For a tree $T=(V_T,E_T,l_T)$, the stretch of an edge $e=uv$ with respect to $T$ is
\[
\str_T(e)
\defeq \frac{l_T(u,v)}{l(e)},
\]
where $l_T(u, v)$ is the length of the unique path between $u$ and $v$ in $T$.
Previous tree embedding algorithms aim to pick a $T$ such that the total
stretch of all edges $e$ in $G$ is small~\cite{AlonKPW95,AbrahamN12}.
A popular alternate goal is to show that the expected stretch of any edge
is small, and these two definitions are closely related~\cite{AlonKPW95,CharikarCGGP98} .
Our other crucial definition is the discounting of high stretches
by adopting the notion of $\ell_p$-stretch:
\[
  \str_{T}^{p}(e) \defeq \left( \str_{T}(e) \right)^{p}.
\]

These two definitional changes greatly simplify the construction of
low stretch embeddings.
It also allows the combination of existing algorithms in a robust manner.
Our algorithm is based on the bottom-up clustering algorithm used to generate
AKPW low-stretch spanning trees~\cite{AlonKPW95}, combined
with the top-down decompositions common in recent
algorithms~\cite{Bartal96,ElkinEST08,AbrahamBN08,AbrahamN12}.
Its guarantees can be stated as follows:

\begin{theorem}
\label{thm:main}
Let $G = (V, E, \length)$ be a weighted graph with $n$ vertices and $m$ edges.
For any parameter $p$ strictly between $0$ and $1$,
we can construct a distribution over trees embeddable in $G$ such that for any edge $e$
its expected $\ell_{p}$-stretch in a tree picked from
this distribution is $\Oh((\frac{1}{1 - p})^2 \log^{p} n)$.
Furthermore, a tree from this distribution can be picked
in expected $\Oh(\frac{1}{1 - p} m \log\log{n})$ time in the RAM model.
\end{theorem}

We will formally define embeddability, as well as other notations,
in Section~\ref{sec:background}.
An overview of our algorithm for generating low $\ell_{p}$-stretch
embeddable trees is in Section~\ref{sec:overview}.
We expand on it using existing low-stretch embedding algorithms
in mostly black-box manners in Section~\ref{sec:bartal}.
Then in Section~\ref{sec:faster} we show a two-stage algorithm
that combines bottom-up and top-down routines that gives our
main result.

Although our algorithm runs in $\Oh(m\log\log{n})$ time, the running
time is in the RAM model, and our algorithm calls a sorting subroutine.
As sorting is used to approximately bucket the edge weights, this
dependency is rather mild.
If all edge lengths are between $1$ and $\Delta$, this process can be done
in $\Oh(m \log (\log \Delta))$ time in the pointer machine model,
which is $\Oh(m \log\log{m})$ when $\Delta \leq m^{\poly(\log{m})}$.
We suspect that there are pointer machine algorithms without even this mild dependence on $\Delta$,
and perhaps even algorithms that improve on the runtime of $\Oh(m\log\log{n})$.
Less speculatively, we also believe that our two-stage approach of
combining bottom-up and top-down schemes can be applied with the decomposition scheme of \cite{AbrahamN12} to
generate actual spanning trees (as opposed to merely embeddable Steiner trees) with low $\ell_p$-stretch.  However,
we do not have a rigorous analysis of this approach, which would presumably require a careful interplay with the
radius-bounding arguments in that paper.

\subsection{Related Works}
\label{subsec:related}

Alon et al.~\cite{AlonKPW95} first proposed the notion of low stretch embeddings
and gave a routine for constructing such trees.
They showed that for any graph, there is a distribution over spanning trees
such that the expected stretch of an edge is $\exp (\Oh ( \sqrt{\log{n}
\log\log{n} }))$.
Subsequently, results with improved expected stretch were
obtained by returning an arbitrary tree metric instead of a spanning tree.
The only requirement on requirement on these tree metrics is that they don't
shorten distances from the original graph, and they may also include extra vertices.
However, in contrast to the objects constructed in this paper, they do
not necessarily fulfill the embeddability property.
Bartal gave trees with expected stretch of $\Oh(\log^2{n})$~\cite{Bartal96},
and $\Oh(\log{n}\log\log{n})$~\cite{Bartal98}.
Optimal trees with $\Oh(\log{n})$ stretches are given by Fakcharoenphol et
al.~\cite{FakcharoenpholRT03}, and are known as the FRT trees.
This guarantee can be written formally as
\begin{align*}
\expct_{T} \left[ \str_{T} (e) \right] \leq \Oh(\log{n}).
\end{align*}

Recent applications to SDD linear system solvers has led to renewed interest
in finding spanning trees with improved stretch over AKPW trees. 
The first LSSTs with $\poly(\log n)$ stretch were given by~Elkin
et al.~\cite{ElkinEST08}.
Their algorithm returns a tree such that the expected stretch of an edge is
$\Oh(\log^{2}n\log\log{n})$, which has subsequently been improved to
$\Oh(\log{n}\log\log{n}(\log\log\log{n})^3)$ by
Abraham et al.~\cite{AbrahamBN08} and
to $\Oh(\log{n}\log\log{n})$ by Abraham and Neiman~\cite{AbrahamN12}.

Notationally our guarantee is almost identical to the expected stretch above
when $p$ is a constant strictly less than $1$:
\begin{align*}
\expct_{T} \left[ \str^{p}_{T} (e) \right] \leq \Oh(\log^{p}{n}).
\end{align*}
The power mean inequality implies that our embedding is
weaker than those with $\ell_1$-stretch bounds.
However, at present, $O(\log n)$ guarantees for $\ell_1$-stretch are {\em not known}--the closest
is the result by Abraham and Neiman~\cite{AbrahamN12},
which is off by a factor of $\log\log{n}$.

Structurally, the AKPW low-stretch spanning trees are constructed in
a bottom-up manner based on repeated clusterings~\cite{AlonKPW95}.
Subsequent methods are based on top down decompositions
starting with the entire graph~\cite{Bartal96}.
Although clusterings are used implicitly in these algorithms, our result
is the first that combines these bottom-up and top-down schemes.

\subsection{Applications}
\label{subsec:applications}

The $\ell_{p}$-stretch embeddable trees constructed in this paper can be
used in all existing frameworks that reduce the size of graphs using 
low-stretch spanning trees.
In Appendix~\ref{sec:embedOk}, we check that the larger graph
with Steiner trees can lead to linear operators close to the graph
Laplacian of the original graph.
It allows us to use these trees in algorithms for solving linear systems in
graph Laplacians, and in turn SDD linear systems.
This analysis also generalizes to other convex norms, which means
that our trees can be used in approximate flow~\cite{LeeS13,Sherman13}
and minimum cut~\cite{Madry10b} algorithms.

Combining our algorithm with the recursive preconditioning framework
by Koutis et al.~\cite{KoutisMP11} leads to an algorithm that runs solves
such a system to constant accuracy in $\Oh(m\log{n})$ time.
They are also crucial for the recent faster solver by
Cohen et al.~\cite{CohenKPPR14:arxiv},
which runs in about $m\log^{1/2}n$ time.
Parallelizations of it can be used can also lead to work-efficient
parallel algorithms for solving SDD linear systems with depth of about $m^{1/3}$~\cite{BlellochGKMPT11}, and in turn for spectral
sparsification~\cite{SpielmanS08,KoutisLP12}.
For these parallel applications, ignoring a suitable fraction of
the edges leads to a simpler algorithm with lower depth.
This variant is discussed in Section \ref{subsec:toss}.
On the other hand, these applications can be further improved
by incorporating the recent polylog depth, nearly-linear work
parallel solver by Peng and Spielman~\cite{PengS13:arxiv}.
Consequently, we omit discussing the best bounds possible with the
hope of a more refined parallel algorithm.

\section{Background}
\label{sec:background}

Before we describe our algorithm, we need to formally specify the simple
embeddability property that our trees satisfy.
The notion used here is the same as the congestion/dilation definition
widely used in routing~\cite{Leighton92:book,LeightonMR94}.
It was used explicitly in earlier works on combinatorial preconditioning~\cite{Vaidya91,Gremban96:thesis},
and is implicit in the more recent algorithms.

Informally, an embedding generalizes the notion of a weighted subgraph in two ways.
First, in an embedding of $H$ into $G$, edges in $H$ may correspond to paths in $G$,
rather than just edges.  Second, $H$ may contain Steiner vertices that can be seen as
``shadow copies'' of vertices in $G$.  Edges in $G$ can be apportioned between different
paths and connect to different Steiner vertices, but their weight must be reduced proportionally.

Formally, an embedding can be viewed as a weighted mapping from
one graph to another.
Splitting an edge will make it lighter, and therefore easier to embed.
However, it will also make it harder to traverse, and therefore longer.
As a result, for embeddings it is convenient to view an edge $e$ by both its
length $l(e)$ and weight $w(e)$, which is the reciprocal of its length:
\[
w(e) \defeq \frac{1}{l(e)}.
\]

A path embedding is then a weighted mapping from the edges of a graph
to paths in another.
Such a mapping from a graph $H=(V_H,E_H,l_H)$ to
a graph  $G=(V_G,E_G,l_G)$ is given by the following three functions:\begin{enumerate}
\item
A mapping from vertices of $H$ to those in $G$, $\pi: V_H \rightarrow V_G$.
\item 
A function from each edge $e_H \in E_H$ to a weighted path of $G$,
denoted by $\mathit{Path}(e_H = x_G y_G)$ that goes from 
$\pi(x_G)$ to $\pi(y_G)$.
\item
We let $W_{\mathit{Path}}(e_H, e_G)$ denote the weight of the edge $e_G$ on path
$\mathit{Path}(e_H)$. This value is zero if $e_G \not\in \mathit{Path}(e_H)$.
\end{enumerate} 
The congestion-dilation notion of embeddability can then be formalized
as follows:
\begin{definition}
\label{def:embeddable}
A graph $H$ is \emph{path embeddable}, or simply \emph{embeddable},  into a graph $G$,
if there exists a path embedding $(\pi,\mathit{Path})$ of $H$ into $G$
such that:
\begin{itemize}
  \item for all edges $e \in E_G$, $\sum_{e_H \in E_H} W_{\mathit{Path}}(e_H, e_G) \leq w_G(e_G)$: congestion is at most one, and
  \item for all edges $e_H \in E_H$, $\sum_{e_G \in \mathit{Path}(e_H)}\frac{1}{W_{\mathit{Path}}(e_H, e_G)} \leq l_H(e) = \frac{1}{w_H(e)}$: dilation is at most one.
\end{itemize}
\end{definition}
Note that since $G$ has no self-loops, the definition precludes
mapping both endpoints of an edge in $H$ to the same point in $G$.
Also note that if $H$ is a subgraph of $G$ such that $l_H(e) \geq l_G(e)$,
setting $\pi$ to be the identity function and $\mathit{Path}(e) = e$
and $W_{Path}(e, e) = w_H(e)$ is one way to certify embeddability.

\section{Overview}
\label{sec:overview}

We now give an overview of our main results.
Our algorithm follows the decomposition scheme taken
by Bartal for generating low stretch embeddings~\cite{Bartal96}.
This scheme partitions the graph repeatedly to form a laminar
decomposition, and then constructs a tree from the laminar decomposition.
However, our algorithm also makes use of spanning trees of the decomposition
itself.  As a result we start with the following alternate definition of Bartal
decompositions where these trees are clearly indicated.

\begin{definition}
\label{def:bartal}
    Let $G = (V, E, l)$ be a connected multigraph.
    We say that a sequence of forests $\BB$, where
    \begin{align*}
        \BB = (B_0, B_1, \ldots, B_t),
    \end{align*}
    is a \emph{Bartal decomposition} of $G$
    if all of the following conditions are satisfied:
    \begin{enumerate}
	\item $B_0$ is a spanning tree of $G$ and $B_t$ is an empty graph.
\label{bartal:startEnd}
        \item For any $i \leq t$, $B_i$ is a subgraph of $G$ in the weighted sense.
\label{bartal:spanning}
        \item For any pair of vertices $u, v$ and level $i < t$, if $u$ and $v$ are in the same connected component of $B_{i + 1}$, then they are in the same connected component of $B_i$.
\label{bartal:laminar}
    \end{enumerate}
\end{definition}


Condition~\ref{bartal:spanning} implies that each of the $B_i$s is
embeddable into $G$.
A strengthening of this condition would require the union of all the $B_i$s 
to be embeddable into $G$.
We will term such decompositions {\em embeddable Bartal decompositions}.

Bartal decompositions correspond to laminar decompositions of the graphs:
if any two vertices $u$ and $v$ are separated by the decomposition in level
$i$, then they are also separated in all levels $j>i$.
If $u$ and $v$ are in the same partition in some level $i$, but are separated
in level $i+1$, we say that $u$ and $v$ are first cut at level $i$.
This definition is useful because if the diameters are decreasing,
the stretch of an edge can be bounded using only information
related to level at which it is first cut.

We will work with bounds on diameters, $\dd = (d_0, \ldots d_t)$.
We say that such a sequence is geometrically decreasing if
there exists some constant $0 < c < 1$ such that $d_{i + 1} \leq c d_i$.
Below we formalize a condition when such sequences can be used
as diameter bounds for a Bartal decomposition.

\begin{definition}
\label{def:diamSeq}
A geometrically decreasing sequence $\dd = (d_0 \ldots d_t)$
bounds the diameter of a Bartal decomposition $\mathcal{B}$ if
for all $0 \leq i \leq t$,
\begin{enumerate}
        \item The diameter of any connected component of $B_i$ is at most $d_i$, and	\label{diam:bound}
	\item any edge $e \in B_i$ has length $l(e) \leq \frac{d_i}{\log n}$. \label{diam:maxLen}
\end{enumerate}
\end{definition}
Given such a sequence, the bound $d_i$ for the level where an edge
is first cut dominates its final stretch.
This motivates us to define the $\ell_p$-stretch of an edge
w.r.t.\ a Bartal decomposition as follows:

\begin{definition}
 \label{def:bartalStretch}
Let $\BB$ be a Bartal decomposition with diameter bounds
bounds $\dd$, and $p$ a parameter such that $p> 0$.
The \emph{$\ell_p$-stretch} with respect to $\BB,\dd$
of an edge $e$ with length $l(e)$ that is first cut at level $i$ is
\[
  \str^p_{\BB, \dd}(e) \defeq \left(\frac{d_i}{l(e)}\right)^{p}.
\]
\end{definition}

In Section~\ref{sec:bartal}, we will check rigorously that it suffices
to generate (not necessarily embeddable) Bartal decompositions for which edges are
expected to have small $\ell_p$-stretch.
We will give more details on these transformations later in
the overview as well.

The decomposition itself will be generated using repeated calls
to variants of probabilistic low-diameter
decomposition routines~\cite{Bartal96}.
Such routines allow one to partition a graph into pieces of
diameter $d$ such that the probability of an edge being cut
is at most $\Oh(\log{n} /d)$.
At a high level, our algorithm first fixes a geometrically decreasing sequence
of diameter bounds, then repeatedly decomposes all the pieces of the graph.
With regular ($\ell_1$) stretch, such routines can be shown to
give expected stretch of about $\log^2{n}$ per edge~\cite{Bartal96},
and most of the follow-up works focused on reducing this factor.
With $\ell_p$-stretch on the other hand, such a trade-off is
sufficient for the optimum bounds when $p$ is a constant
bounded away from $1$.

\begin{lemma}
\label{lem:goodProb}
Let $\mathcal{B}$ be a distribution over Bartal decompositions.
If $\dd$ is a geometrically decreasing sequence that bounds the diameter
of any $\BB \in \mathcal{B}$, and
the probability of an edge with length $l(e)$ being
cut on level $i$ of some $\BB \in \mathcal{B}$ is
\[
\Oh\left(\left(\frac{l(e) \log n}{d_i}\right)^{q}\right)
\]
for some $0 < q < 1$.
Then for any $p$ such that $0 < p < q$, we have
\[
\expct_{\BB \in \mathcal{B}}\left[\str_{\BB,\dd}^{p}(e)\right]
\leq \Oh\left(\frac{1}{q - p} \log^{p}{n}\right)
\]
\end{lemma}

Its proof relies on the following fact about geometric series,
which plays a crucial role in all of our analyses.

\begin{fact}
\label{fact:geoSeries}
There is an absolute constant $c_{\mit{geo}}$ such that
if $c$ and $\epsilon$ are parameters such that $c \in [e, e^{2}]$
and $\epsilon > 0$
\[
  \sum_{i = 0}^{\infty} c^{-i\epsilon} =  c_{\mit{geo}} \epsilon^{-1}.
\]
\end{fact}

\Proof
Since $0<c^{-1} < 1$, the sum converges, and equals
\[
\frac{1}{1 - c^{-\epsilon}}
= \frac{1}{1 - \exp(-\epsilon \ln{c})}.
\]
Therefore it remains to lower bound the denominator.
If $\epsilon\geq 1/4$, then the denominator
can be bounded by a constant.
Otherwise, $\epsilon \ln{c} \leq 1/2$, and we can invoke
the fact that $\exp(-t)  \leq 1 - t/2$ when $t \leq 1/2$ to obtain
\[
1 - \exp(-\epsilon \ln{c})
\geq \epsilon \ln{c}.
\]
Substituting in the bound on $c$ and
this lower bound into the denominator then gives the result.
\QED

\Proofof{Lemma~\ref{lem:goodProb}}
If an edge is cut at a level with $d_i \leq l(e)\log n$, its stretch is at most
$\log n$, giving an $\ell_p$-stretch of at most $\log^{p} n$.
It remains only to consider the levels with $d_i \geq l(e)\log n$.
Substituting the bounds of an edge cut on level $i$ and the probability of it being cut
into the definition of $\ell_p$-stretch gives:
\[
\expct_{\BB}\left[ \str_{\BB,\dd}^{p}\left(e\right) \right]
\leq \sum_{i, d_i \geq \log{n} l(e)} \left(\frac{d_i}{l(e)}\right)^{p} \Oh\left( \left( \frac{l(e) \log{n}}{d_i} \right)^{q} \right)\\
= \Oh \left( \log^{p} {n} \sum_{i, d_i \geq \log{n} l(e)} \left(\frac{l(e) \log{n} }{d_i}\right)^{q - p}\right).
\]
Since an edge $e$ is only cut in levels where
$d_i \geq l(e) \log{n}$ and the $d_i$s are geometrically increasing,
this can be bounded by
\[
\Oh \left( \log^{p}{n} \sum_{i = 0} c^{-i(q  - p)} \right)
\]
Invoking Fact~\ref{fact:geoSeries} then gives a bound of
$\Oh \left(\frac{1}{q - p} \log^{p}{n} \right)$.
\QED

This is our approach for showing that a Bartal decomposition
has small $\ell_p$-stretch, and it remains to convert
them into embeddable trees.
This conversion is done in two steps: we first show
how to obtain a decomposition such that all of the
$B_i$s are embeddable into $G$, and then we give
an algorithm for converting such a decomposition
into a Steiner tree.
To accomplish the former, we first ensure that each $B_i$
is embeddable by choosing them to be subgraphs.
Then we present pre-processing and post-processing procedures
that converts such a guarantee into embeddability
of all the $B_i$s simultaneously.

\begin{figure}[t!]
  \begin{center}
    \includegraphics{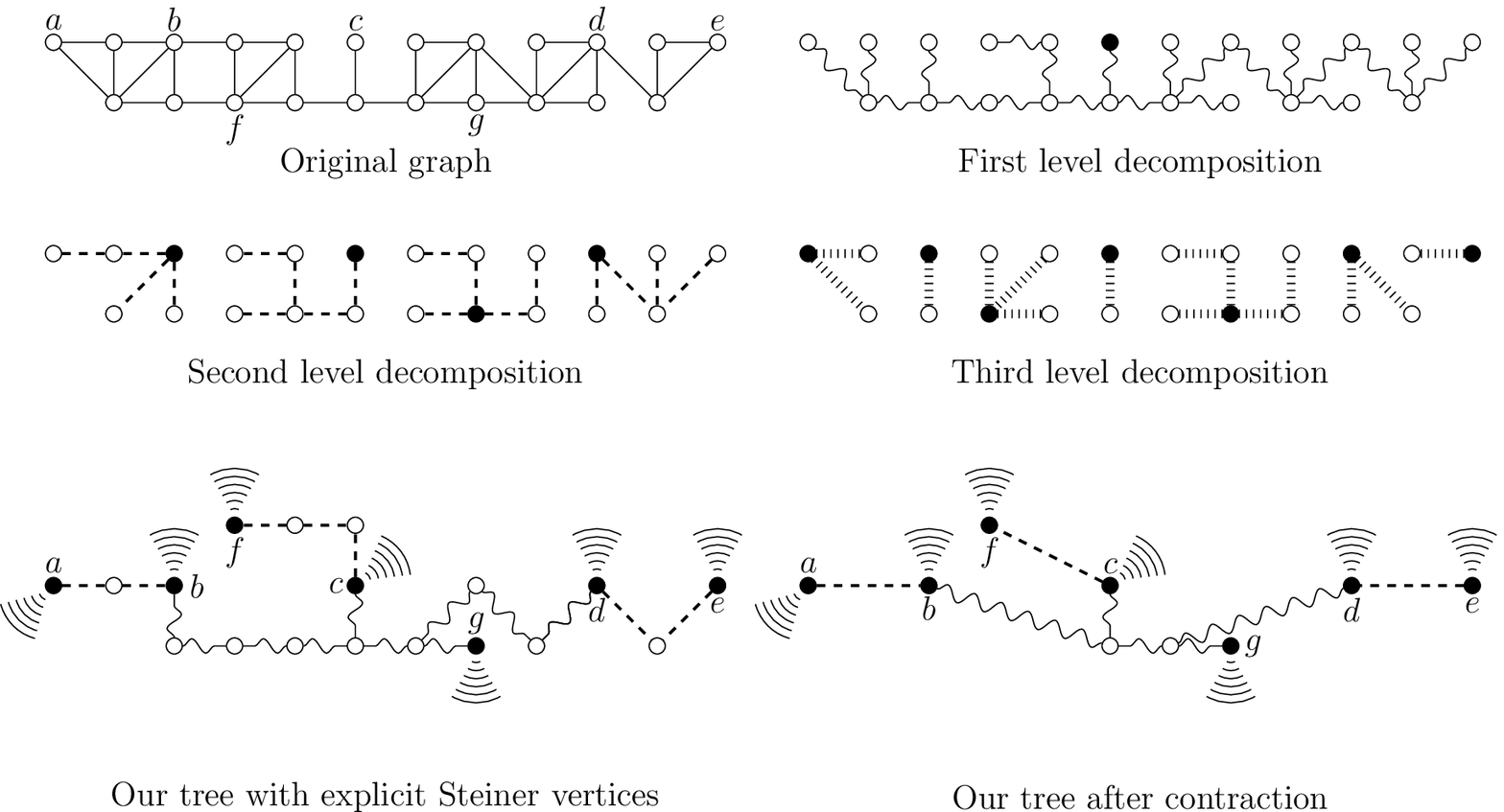}
  \end{center}
  \caption{Bartal decomposition and the tree produced for a particular graph}
  \label{fig:buildTreeFig}
\end{figure}

In order to obtain a tree from the decomposition,
we treat each cluster in the laminar decomposition
as a Steiner vertex, and join them using parts of $B_i$s.
This step is similar to Bartal trees in that it identifies
centers for each of the $B_i$s, and connects the centers
between one level and the next.
However, the need for the final tree to be embeddable means
that we cannot use the star-topology from Bartal trees~\cite{Bartal96}.
Instead, we must use part of the $B_i$s between the centers.
As each $B_i$ is a forest with up to $n$ edges, a tree obtained
as such may have a much larger number of Steiner vertices.
As a result, the final step involves reducing the size of this tree
by contracting the paths connecting the centers. 
This process is illustrated in Figure~\ref{fig:buildTreeFig}.

In Section~\ref{sec:bartal}, we give the details on these steps that
converts Bartal decompositions to embeddable trees.
Furthermore, we check that Bartal's algorithm for generating
such trees meets the good cutting probability requirements of
Lemma~\ref{lem:goodProb}.
This then gives the following result:

\begin{restatable}{lemma}{bartalAlgo}
\label{lem:bartalAlgo}
Given a graph $G$ with weights are between $[1, \Delta]$,
for the diameter sequence
$\dd$ where $d_0 = 2 n \Delta, d_1 = 2^{-1} n \Delta, \ldots
d_t < 1$,
we can create a distribution over Bartal decompositions
with diameters bounded by $\dd$ such that for any edge $e$
and any parameter $0 < p < 1$,
\[
\expct_{\BB}\left[\str_{\BB,\dd}^{p} \left(e\right)\right]
\leq \Oh\left( \frac{1}{1 - p} \log^{p} n \right).
\]
Furthermore, a random decomposition from this distribution
can be sampled with high probability in
$\Oh(m \log(n\Delta) \log{n})$ time in the RAM model.
\end{restatable}

This routine plus the transformations gives a simple algorithm
for constructing low $\ell_p$-stretch embeddable trees.
with expected stretch matching the bound stated our
main result, Theorem~\ref{thm:main}.
However, the running time of $\Oh(m \log(n\Delta) \log{n})$ is
more than the current best for finding low-stretch spanning
trees~\cite{AbrahamN12}, as well as the $\Oh(m\log^2{n})$ running
time for finding Bartal trees.

Our starting point towards a faster algorithm is the difference between
our simplified routine and Bartal's algorithm.
Bartal's algorithm, as well as subsequent algorithms~\cite{ElkinEST08}
ensure that an edge participates in only $\Oh(\log{n})$ partitions.
At each step, they work on a graph obtained by contracting
all edges whose lengths are less than $d_{i}/\poly(n)$.
This coupled with the upper bound of edge lengths from
Definition~\ref{def:diamSeq},~Part~\ref{diam:maxLen}
and the geometric decrease in diameter bounds
gives that each edge is involved in $\Oh(\log(\poly(n))) = \Oh(\log{n})$
steps of the partition.

As a path in the tree has at most $n$ edges,
the additive increase in stretch caused by these
shrunken edges is negligible.
Furthermore, the fact that the diameter that we partition upon
decreases means that once we uncontract an edge, it remains
uncontracted in all future steps.
Therefore, these algorithms can start from the initial contraction
for $d_0$, and maintain all contractions in work proportional to
their total sizes.

When viewed by itself, this contraction scheme is almost identical
to Kruskal's algorithm for building minimum spanning trees (MSTs).
This suggests that the contraction sequence can be viewed as
another tree underlying the top-down decomposition algorithm.
This view also leads to the question of whether other trees can
be used in place of the MST.
In Section~\ref{sec:faster}, we show that if the AKPW low-stretch
spanning tree is used instead, each edge is expected to participate
in $\Oh(\log\log{n})$ levels of the top-down decomposition scheme.
Combining this with a $\Oh(m\log\log{n})$ time routine in the RAM
model for finding the AKPW low-stretch spanning tree and a faster
decomposition routine then leads to our faster algorithm.

Using these spanning trees to contract parts of the graph
leads to additional difficulties in the post-processing steps
where we return embeddable Steiner trees.
A single vertex in the contracted graph may correspond to
a large cluster in the original graph.
As a result, edges incident to it in the decomposition may need
to be connected by long paths.
Furthermore, the total size of these paths may be large,
which means that they need to be treated implicitly.
In Section~\ref{subsec:iCanHazTree}, we leverage the tree
structure of the contraction to implicitly compute the reduced tree.
Combining it with the faster algorithm for generating Bartal
decompositions leads to our final result as stated in Theorem~\ref{thm:main}.


\section{From Bartal Decompositions to Embeddable Trees}
\label{sec:bartal}

In this section, we show that embeddable trees can be obtained
from Bartal decompositions
using the process illustrated in Figure~\ref{fig:buildTreeFig}.
We do this in three steps: exhibiting Bartal's algorithm in
Section~\ref{subsec:bartal}, showing that a decomposition routine
that makes each $B_i$ embeddable leads to a routine that generates
embeddable decompositions in Section~\ref{subsec:embed},
and giving an algorithm for finding a tree from the decomposition
in Section~\ref{subsec:makeTree}.
We start by formally describing Bartal's algorithm for decomposing
the graph.


\subsection{Bartal's Algorithm}
\label{subsec:bartal}

Bartal's algorithm in its simplest form can be viewed as repeatedly
decomposing the graph so the pieces have the diameter guarantees
specified by $\dd$.
At each step, it calls a low-diameter probabilistic decomposition routine
with the following guarantees.

\begin{lemma}[Probabilistic Decomposition]
\label{lem:partition}
	There is an algorithm \textsc{Partition} that given a graph $G$ with $n$ vertices and $m$ edges, and a diameter parameter $d$,
returns a partition of $V$ into
$V_1 \disjunion V_2  \disjunion \ldots  \disjunion V_k$ such that:
\begin{enumerate}
	\item The diameter of the subgraph induced on each $V_i$ is at most $d$ with
    high probability, certified by a shortest path tree on $V_i$ with diameter
    $d$, and
	\item for any edge $e = uv$ with length $l(e)$, the probability that
$u$ and $v$ belong to different pieces is at most $\Oh(\frac{l(e) \log{n}}{d})$.
\end{enumerate}

   Furthermore, \textsc{Partition} can be implemented using
 one call to finding a single source shortest path tree
on the same graph with all vertices connected to
a super-source by edges of length between $0$ and $d$.
\end{lemma}

This routine was first introduced by Bartal to construct these
decompositions.
It and the low diameter decompositions that it's based on
constructed each $V_i$ in an iterative fashion.
Miller et al.~\cite{MillerPX13} showed that a similar procedure can be viewed
globally, leading to the implementation-independent view
described above.
Dijkstra's algorithm (Chapter 24 of~\cite{CormenLRS01:book}) then allows one to obtain a running time of $\Oh((m  + n) \log{n})$.
It can be further sped up to $\Oh(m + n\log{n})$
using Fibonacci heaps due to Fredman and Tarjan~\cite{FredmanT87},
and to $\Oh(m)$ in the RAM model by Thorup~\cite{Thorup00}.
In this setting where approximate answers suffice, a running time of
$\Oh(m + n \log\log{\Delta})$ was also obtained by Koutis et al.~\cite{KoutisMP11}.
As our faster algorithm only relies on the shortest paths algorithm in a more
restricted setting, we will use the most basic $\Oh(m\log{n})$
bound for simplicity.

We can then obtain Bartal decompositions by invoking this 
routine recursively.
Pseudocode of the algorithm is given in Figure~\ref{fig:bartal}.
The output of this algorithm for a suitable diameter sequence
gives us the decomposition stated in Lemma~\ref{lem:bartalAlgo}.

\begin{figure}

\vskip 0.2in
\noindent
\centering
\fbox{
\begin{minipage}{6in}
\noindent $\BB = \textsc{DecomposeSimple} (G, \dd)$,
where  $G$ is a multigraph, and $\dd$ are diameter bounds.
\begin{enumerate}
	\item Initialize $\BB$ by setting $B_0$ to
a shortest path tree from an arbitrary vertex in $V_G$.
	\item For $i = 1 \ldots t$ do
		\begin{enumerate}
			\item Initialize $B_i$ to empty.
			\item Remove all edges $e$ with $l(e) \geq \frac{d_{i}}{\log{n}}$ from $G$.
			\item For each subgraph $H$ of $G$ induced by a connected component of
        $B_{i-1}$ do
				\begin{enumerate}
					\item $G_1 \ldots G_k \leftarrow \textsc{Partition}(H, d_i)$.
					\item Add shortest path trees in each $G_j$ to $B_i$.
				\end{enumerate}
		\end{enumerate}
	\item Return $\BB$.
\end{enumerate}
\end{minipage}
}
\vskip 0.2in
\caption{Bartal's Decomposition Algorithm}
\label{fig:bartal}

\end{figure}




\Proofof{Lemma~\ref{lem:bartalAlgo}}
Consider the Bartal distributions produced by
running $\textsc{DecomposeSimple}(G,\dd)$.
With high probability it returns a decomposition $\BB$.
We first check that $\BB$ is a Bartal decomposition.
Each tree in $B_i$s is a shortest path tree on a cluster of vertices
formed by the partition.
As these clusters are disjoint, $B_i$ is a subgraph in the weighted sense.
Since the algorithm only refines partitions,  once two vertices
are separated, they remain separated for any further partitions.
Also, the fact that $B_0$ is spanning follows from the
initialization step, and $B_t$ cannot contain any edge since
any edge has length at least $1$ and $d_t < 1$.

We now show that $\dd$ are valid diameter bounds for any decomposition
produced with high probability.
The diameter bounds on $d_i$ follow from the guarantees of
\textsc{Partition} and the initialization step.
The initialization of $d_0 = 2 n \Delta$ also ensures that no edge's
length is more than $\frac{d}{\log{n}}$, and this invariant is kept
by discarding all edges longer than $\frac{d}{c \log{n}}$ before
each call to \textsc{Partition}.

The running time of the algorithm follows from $t \leq \Oh(\log(n \Delta))$
and the cost of the shortest path computations at all the steps.
It remains to bound the expected $\ell_p$-stretch of an edge
$e$ w.r.t.\ the decomposition.
When $l(e) \geq \frac{d_i}{c \log{n}}$, a suitable choice of constants
allows us to bound the probability of $e$ being cut by $1$.
Otherwise, $e$ will not be removed unless it is already cut.
In case that it is in the graph passed onto $\textsc{Partition}$, the
probability then follows from Lemma~\ref{lem:partition}.
Hence the cutting probability of edges satisfies Lemma~\ref{lem:goodProb},
which gives us the bound on stretch.
\QED

\subsection{Embeddability by Switching Moments}
\label{subsec:embed}

We now describe how to construct \emph{embeddable} Bartal decomposition
by using a routine that returns Bartal decompositions.
This is done in three steps:
pre-processing the graph to transform the edge lengths of $G$ to form
$G'$, running the decomposition routine on $G'$ for a different parameter
$q$, and post-processing its output.

\begin{figure}[ht]
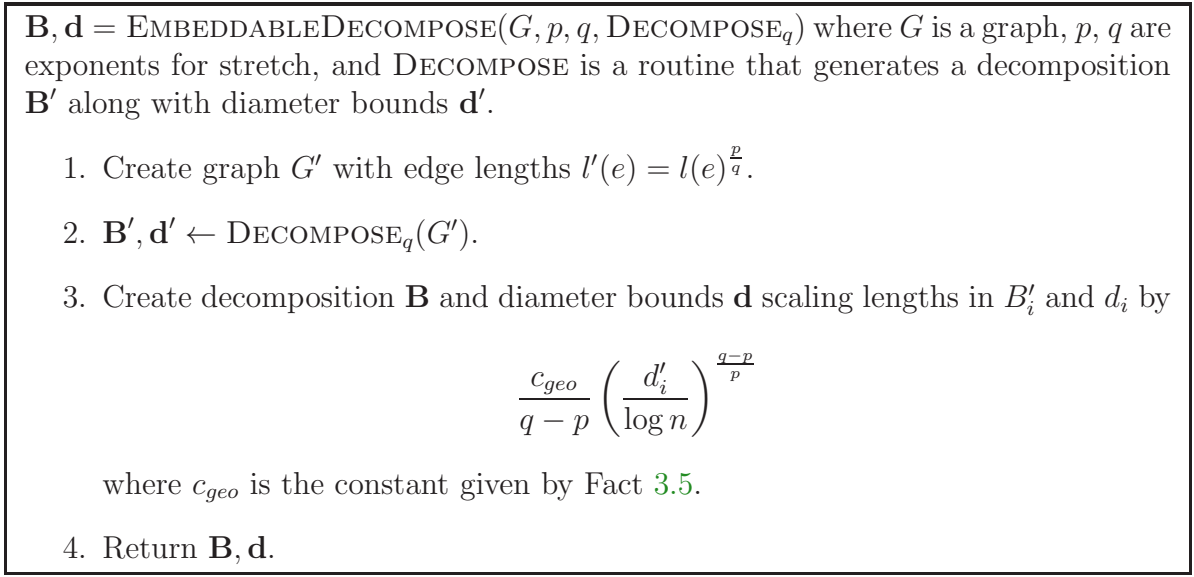

\vskip 0.2in
\centering
\fbox{
\begin{minipage}{6in}
\noindent $\BB, \dd = \textsc{EmbeddableDecompose} (G, p, q, \textsc{Decompose}_q)$ where $G$ is a graph, $p$, $q$ are exponents
for stretch, and $\textsc{Decompose}$ is a routine that
generates a decomposition $\BB'$ along with diameter bounds $\dd'$.
\begin{enumerate}
	\item Create graph $G'$ with edge lengths $l'(e) = l(e)^{\frac{p}{q}}$.\label{line:pre}
	\item $\BB', \dd' \leftarrow \textsc{Decompose}_q(G')$.
	\item Create decomposition $\BB$ and diameter bounds $\dd$
scaling lengths in $B'_i$ and $d_i$ by
\[
\frac{c_{geo}}{q - p} \left( \frac{d_i'}{\log{n}} \right)^{\frac{q - p}{p}} 
\]
where $c_{geo}$ is the constant given by Fact~\ref{fact:geoSeries}.
\label{line:post}
	\item Return $\BB, \dd$.
\end{enumerate}
\end{minipage}
}
\caption{Using a generic decomposition routine to generate an embeddable decomposition}
\label{fig:embeddableDecompose}
\end{figure}

Pseudocode of this conversation procedure is given in
Figure~\ref{fig:embeddableDecompose}.
Both the pre-processing and post-processing steps are
deterministic, linear mappings.
As a result, we can focus on bounding the expected stretch
of an edge in the decomposition given by \textsc{Decompose}.

We first verify that $\dd$ is a geometrically decreasing sequence
bounding the diameters of $\BB$.

\begin{lemma}
\label{lem:diamBoundsOk}
If $\BB'$ is a Bartal decomposition of $G'$ whose diameters are
bounded by $\dd'$, then $\dd$ is geometrically decreasing sequence
that bound the diameter of $\BB$.
\end{lemma}

\Proof

The post-processing step scales the difference between adjacent
$d_i'$s by an exponent of $\frac{q}{p}$, which is at least $1$
since $q > p$.
Therefore $\dd$ is also a geometrically decreasing sequence.
As the lengths in $B_i'$ and $d_i'$ are scaled by the same factor,
$d_i$ remains an upper bound for the diameter of $B_i$.
Also, since $d'_i \geq l'(e)\log{n} = l(e)^{\frac{p}{q}}\log{n} $, we have
\[
d_i
=
\frac{c_{geo}}{q - p} \left( \frac{d_i'}{\log{n}} \right)^{\frac{q - p}{p}} d'_i
\ge l(e)\log{n}.
\]
Therefore $\dd$ upper bounds the diameters of $\BB$ as well.

\QED

We now check that $\BB$ is a subgraph in the weighted case, which
makes it an embeddable Bartal decomposition.

\begin{lemma}
\label{lem:subgraph}
For any edge $e$ we have
\[
\sum_{i} w_{B_i}(e) \leq w(e).
\]
\end{lemma}

\Proof
Combining the pre-processing and post-processing steps gives that the
total weight of $e$ in all the layers is:
\[
\sum_{i} w_{B_i}(e)
= \sum_{i, e \in B_i} \frac{1}{l_{B_i}(e)}
=
\frac{p - q}{c_{geo}} \sum_{i, e \in B_i} \left( \frac{\log{n}}{d_i'} \right)
    ^{\frac{q - p}{p}}  w(e)^{\frac {p}{q}}.
\]
Showing that this is at most $w(e)$ is therefore equivalent to showing
\[
\frac{p - q}{c_{geo}}  \sum_{i, e \in B_i}  \left( \frac{\log{n}}{d_i' w(e)^{\frac{p}{q}}} \right)^{\frac{q - p}{p}} \leq 1.
\]
Here we make use of the condition that the levels in which edge $e$
appears have $d_i' \geq l'(e) \log{n}$.
Substituting in $l'(e) = w(e)^{-\frac{p}{q}}$ into this bound on $d_i'$ gives:
\begin{align*}
d_i'
&\geq w(e)^{-\frac{p}{q}} \log{n}\\
d_i' w(e)^{\frac{p}{q}}
& \geq \log{n}.
\end{align*}
As $d'_i$s are decreasing geometrically, this means that these terms
are a geometrically decreasing sequence whose first term can be bounded
by $1$.
Fact~\ref{fact:geoSeries} then gives that the summation is bounded by
$\frac{c_{geo} p }{q - p} \leq \frac{c_{geo}}{q - p}$, which cancels
with the coefficient in front of it.
\QED

We can also check that the stretch of an edge $e$ w.r.t.\ $\BB, \dd$
is comparable to its stretch in $\BB', \dd'$.

\begin{lemma}
\label{lem:stretchpq}
For parameters $0 < p < q < 1$, 
the $\ell_p$-stretch of an edge $e$ in $G$ w.r.t.\ $\BB,\dd$ and its
$\ell_q$-stretch in $G'$ w.r.t.\ $\BB', \dd'$ are related by
\[
        \str^p_{\BB,\dd}(e) = \Oh\left(\frac{1}{q - p} \log^{p - q} n \cdot \str^{q}_{\BB',\dd'}(e)\right).
\]
\end{lemma}

\Proof
Rearranging scaling on $d'_i$ used to obtain $d_i$ gives
\[
d_i = \frac{c_{geo}}{q - p} \left( \frac{d_i'}{\log{n}} \right)^{\frac{q - p}{p}} d'_i
= \frac{c_{geo}}{q - p} \log^{\frac{p-q}{p}}{n} \cdot {d'_i}^{\frac{q}{p}}.
\]
We can then relate the stretch of an edge in the new decomposition
with that of its $\ell_q$-stretch in $G'$.
For an edge cut at level $i$, we have
\[
\str_{\BB, \dd}(e)
= \frac{d_i}{l(e)}
=  \frac{c_{geo}}{q - p} \log^{\frac{p - q}{p}}{n} \frac{{d'_i}^{\frac{q}{p}}}{l'(e)^{\frac{q}{p}}}.
=  \frac{c_{geo}}{q - p}  \log^{\frac{p - q}{p}}{n} \left(\str^{q}_{\BB', \dd'}(e)\right)^{\frac{1}{p}}.
\]
Taking both sides to the $p$-th power, and using the fact that
$p < 1$, then gives the desired bound.
\QED

It's worth noting that when $p$ and $q$ are bounded away from $1$ by
constants, this procedure is likely optimal up to constants.
This is because the best $\ell_p$-stretch that one could obtain in these
settings are $\Oh(\log^{p}{n})$ and $\Oh(\log^{q}{n})$ respectively.

\subsection{From Decompositions to Trees}
\label{subsec:makeTree}

It remains to show that an embeddable decomposition can be
converted into an embeddable tree.
Our conversion routine is based on the laminar-decomposition
view of the decomposition.
From the bottommost level upwards, we iteratively reduce the
interaction of each cluster with other clusters to a single vertex in it,
which we term the centers.
Centers can be picked arbitrarily,
but to enforce the laminar decomposition view,
we require that if a vertex $u$ is a center on level $i$, it is also
a center on level $i + 1$ and therefore all levels $j > i$.
Once the centers are picked, we can connect the clusters starting
at the bottom level, by connecting all centers of level $i + 1$ to
the center of the connected component they belong to at level $i$.
This is done by taking the part of $B_i$ involving these centers.
We first show that the tree needed to connect them has size
at most twice the number of centers.

\begin{fact}
\label{fact:smallerTree}
Given a tree $T$ and a set of $k$ vertices $S$,
there is a tree $T_S$ on $2k - 1$
vertices including these $k$ leaves such that:
\begin{itemize}
	\item The distances between vertices in $S$ are the same in $T$ and $T_S$.
	\item $T_S$ is embeddable into $T$.
\end{itemize}
\end{fact}

\Proof
The proof is by induction on the number of vertices in $T$.
The base case is when $T$ has fewer than $2k - 1$ vertices,
where it suffices to set $T_S = T$.
For the inductive case suppose the result is true for all trees
with $n$ vertices, and $T$ has $n + 1$ vertices.
We will show that there is a tree $T'$ on $n$ vertices that preserves
all distances between vertices in $S$, and is embeddable into $T$.

If $T$ has a leaf that's not in $S$, removing it and the edge
incident to it does not affect the distances between the vertices
in $S$, and the resulting tree $T'$ is a subgraph and therefore embeddable
into $T$.
Otherwise, we can check via a counting argument that there is a vertex
$u$ of degree $2$ that's not in $S$.
Let this vertex and its two neighbors be $u$ and $v_1$, $v_2$ respectively.
Removing $u$ and adding an edge between $v_1v_2$ with weight
$l(uv_1)+l(uv_2)$ preserves distances.
This new tree $T'$ is embeddable in $T$ by mapping $v_1v_2$ to the path
$v_1-u-v_2$ with weights equaling the weights of the two edges.

Since $T'$ has $n$ vertices, the inductive hypothesis gives the existence
of a tree $T_S$ meeting the requirements.
As $T'$ is embeddable into $T$, $T_S$ is embeddable into $T$ as well.
\QED

Invoking this routine repeatedly on the clusters then leads to
the overall tree.
Pseudocode of this tree construction algorithm is given
in Figure~\ref{fig:buildTree}.

\begin{figure}[ht]
\vskip 0.2in
\centering
\fbox{
\begin{minipage}{6in}
\noindent $T = \textsc{buildTree} (G, \BB)$ where $\BB$ is a Bartal decomposition of $G$.
\begin{enumerate}
	\item Designate a center vertex for each connected component of each level of
    $\BB$ such that if $u$ is a center vertex on level $i$, it is also a
    center vertex on level $i + 1$.
	\item For each connected component on level $i$.
		\begin{enumerate}
			\item Find all center vertices in level $i + 1$ contained in this piece. 
      \item Connect these vertices using the small sized equivalent of $B_i$ given by Fact~\ref{fact:smallerTree}.
		\end{enumerate}
	\item Return $T$.
\end{enumerate}
\end{minipage}
}
\caption{Constructing a Steiner tree from a Bartal decomposition}
\label{fig:buildTree}
\end{figure}

\begin{lemma}
\label{lem:buildTree}
Given a graph $G$ and an embeddable Bartal decomposition $\BB$,
\textsc{BuildTree} gives an embeddable tree $T$ with $\Oh(n)$ vertices
containing $V$ such that for any geometrically decreasing sequence
$\dd$ that bounds the diameters of $\BB$ and any edge $e$ we have
\[
\str_{T}(e) = \Oh(\str_{\BB, \dd}(e)).
\]
\end{lemma}

\Proof
We first bound the total size of $T$.
Note that the number of vertices added is proportional to the decrease
in number of components.
Since the initial number of clusters is $n$, $T$ has at most $2n - 1$ vertices.

Fact~\ref{fact:smallerTree} gives that the trees used to
connect the level $i + 1$ clusters are embeddable into the
corresponding connected component of $B_i$.
Since the vertices in these clusters are disjoint and 
$\cup_i B_i$ is embeddable into $G$,
$T$ is also embeddable into $G$.

It remains to bound the stretch of edges w.r.t.\ $T$.
For an edge $e = uv$ that's cut at level $i$, consider the the path
from $u$ to the centers of the clusters levels $t, t - 1, \ldots i$.
The diameter bounds give that the distance traversed on level
$j$ is bounded by $d_j$.
As $\dd$ is a geometrically decreasing sequence, the total length
of this path is bounded by $\Oh(d_i)$.
A similar argument can be applied to $v$, and since $u$ and $v$
is cut at level $i$, the centers on level $i$ are the same.
Therefore, the distance between $u$ and $v$ in the tree can
be bounded by $\Oh(d_i)$, giving the bound on stretch.
\QED

Combining these pieces leads to an algorithm
generating low $\ell_p$-stretch embeddable trees.

\begin{lemma}
\label{lem:bartalGood}
Let $G = (V, E, \weight)$ be a weighted graph with
$n$ vertices and $m$ edges and weights $w:E\to[1, \Delta]$,
and $p$ be any parameter strictly between $0$ and $1$.
We can construct a distribution over Bartal decompositions
such that for any edge $e$, its expected $\ell_{p}$-stretch
in a decomposition picked from this distribution is
$\Oh((\frac{1}{1 - p})^2\log^{p} n)$.
\end{lemma}

\Proof

Consider running \textsc{EmbeddableDecompose} with $q = \frac{1 + p}{2}$,
and \textsc{DecomposeSimple} with the parameters given by
Lemma~\ref{lem:bartalAlgo} as the decomposition procedure.
By Lemmas~\ref{lem:bartalAlgo}~and~\ref{lem:stretchpq},
the expected stretch of an edge $e$ in the post-processed
decomposition $\BB$ w.r.t. diameter bounds $\dd$ is:
\[
\Oh \left(\frac{1}{q - p}  \log^{p - q} \frac{1}{1 - q} \log^{q} n\right)
= \Oh \left( \left(\frac{1}{1 - p}\right)^2 \log^{p} n\right).
\]
Running \textsc{BuildTree} on this decomposition then gives a tree
where the expected stretch of edges are the same.
The embeddability of this tree also follows from the embeddability of
$\BB$ given by Lemma~\ref{lem:subgraph}.

To bound the runing time, note that as $0 \leq \frac{p}{q} < 1$,
the lengths of edges in the pre-processed graph $G'$ are also
between $1$ and $\Delta$.
Both the pre and post processing steps consist of only
rescaling edge weights, and therefore take linear time.
The total running time then follows from Lemma~\ref{lem:bartalAlgo}.
\QED

\section{Two-Stage Tree Construction}
\label{sec:faster}

We now give a faster algorithm for constructing Bartal decompositions.
The algorithm proceeds in two stages.
We first quickly build a lower quality decomposition using the same scheme
as the AKPW low stretch spanning tree~\cite{AlonKPW95}.
Then we proceed in the same way as Bartal's algorithm and refine the
decompositions in a top-down manner.
However, with the first stage decomposition, we are able to construct a Bartal
decomposition much faster.

Both the AKPW decomposition and the way that our Bartal decomposition
routine uses it relies on repeated clustering of vertices.
Of course, in an implementation, such clusterings will be represented
using various linked-list structures.
However, from an analysis perspective, it is helpful to view them as quotient graphs.
For a graph $G$ and a subset of edges $A$, we let the quotient graph $G / A$
be the graph formed by the connected components of $A$.
Each of these components corresponding to subsets of vertices becomes
a single vertex in $G / A$, and the edges have their vertices relabeled accordingly.
For our algorithms, it is essential for us to keep multi-edges as separate copies.
As a result, all the graphs that we deal with in this section are potentially
multi-graphs, and we will omit this distinction for simplicity.

The main advantages offered by the AKPW decomposition are
\begin{itemize}
	\item it is a bottom-up routine that can be performed in linear time, and
	\item each edge only participates in $\Oh(\log\log{n})$ steps of the refinement process in expectation, and
	\item all partition routines are done on graphs with diameter $\poly(\log n)$.
\end{itemize}
The interaction between the bottom-up AKPW decomposition scheme
and the top-down Bartal decomposition leads to some distortions.
The rest of this section can be viewed as analyzing this distortion,
and the algorithmic gains from having it.
We will show that for an appropriately constructed AKPW decomposition,
the probability of an edge being cut can be related to a quantity in
the $\ell_q$ norm for some $p < q < 1$.
The difference between these two norms then allows us to absorb
distortions of size up to $\poly{\log{n}}$, and therefore not affecting
the quality of the resulting tree.
Thus we will work mostly with a different exponent $q$ in this
section, and only bring things back to an exponent in $p$ at the very end.

Both the AKPW and the top-down routines will issue multiple calls to $\textsc{Partition}$.
In both cases the granularity of the edge weights will be $\poly(\log n)$.
As stated in Section 3, $\textsc{Partition}$ can be implemented in linear time in the RAM model,
   using the rather involved algorithm presented in~\cite{Thorup00}.
In practice, it is also possible to use the low granularity of edge weights and use Dial's algorithm~\cite{Dial69},
worsening the total running time of our algorithm to $\Oh(m \log \log n + \log \Delta \ \poly(\log n))$ when all edge lengths are in the range $[1, \Delta]$.
Alternatively, we can use the weight-sensitive shortest path algorithm from~\cite{KoutisMP11},
which works in the pointer machine model, but would be slower by a factor of $\Oh(\log \log \log n)$.

\subsection{The AKPW Decomposition Routine}

\begin{figure}[ht]
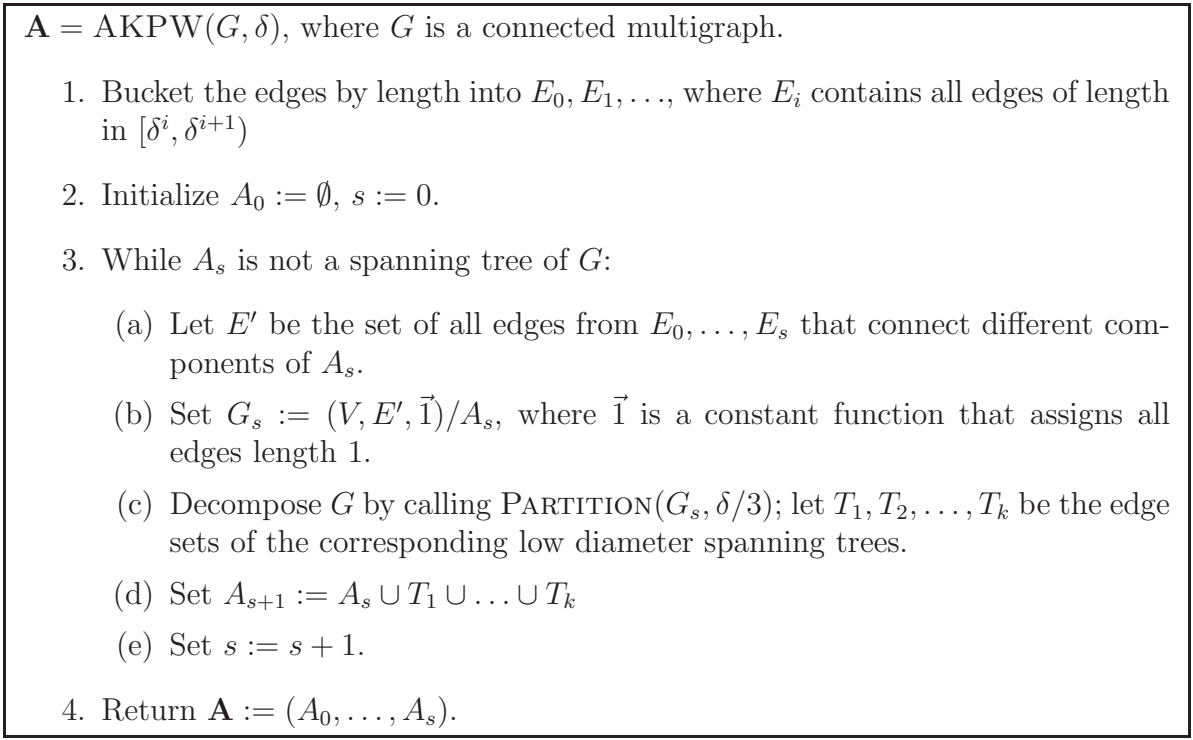

\noindent
\centering
\fbox{
\begin{minipage}{6in}
\noindent $\AA = \textsc{AKPW} (G, \delta)$,
where $G$ is a connected multigraph.
\begin{enumerate}
\item Bucket the edges by length into $E_0, E_1, \ldots$, where $E_i$ contains all edges of length in $[\delta^{i}, \delta^{i + 1})$
\item Initialize
    $A_0 := \emptyset$,
    $s := 0$.
\item While $A_s$ is not a spanning tree of $G$:
\label{line:terminate}
    \begin{enumerate}
        \item Let $E'$ be the set of all edges from $E_0, \ldots, E_s$ that connect different components of $A_s$.
        \item Set $G_s := (V, E', \ones) / A_s$,
              where $\ones$ is a constant function that assigns all edges length $1$.
        \item Decompose $G$ by calling $\textsc{Partition}(G_s, \delta / 3)$; let $T_1, T_2, \ldots, T_k$ be the edge sets of the corresponding low diameter spanning trees.

        \item Set $A_{s + 1} := A_s \cup T_1 \cup \ldots \cup T_k$
        \item Set $s := s + 1$.
    \end{enumerate}
\item Return $\AA := (A_0, \ldots, A_s)$.
\end{enumerate}
\end{minipage}
}
\caption{The routine for generating AKPW decompositions}
\label{fig:akpw}
\end{figure}

We first describe the AKPW algorithm for generating decomposition.
The decomposition produced is similar to Bartal decompositions,
although we will not impose the strict conditions on diameters in our definition.
\begin{definition}
\label{def:akpw}
    Let $G = (V, E, l)$ be a connected multigraph.
    We say that a sequence of forests $\AA$, where
    \[
        \AA = (A_0, A_1, \ldots, A_s),
    \]
    is an \emph{AKPW decomposition} of $G$ with parameter $\delta$ if:
    \begin{enumerate}
        \item $A_s$ is a spanning tree of $G$.
		\item For any $i < t$, $A_i \subseteq A_{i + 1}$.
		\item The diameter of each connected component in $A_i$ is at most $\delta^{i + 1}$.
	\end{enumerate}
\end{definition}
Pseudocode for generating this decomposition is given in Figure~\ref{fig:akpw}.
We first bound the diameters of each piece, and the probability of an edge being cut in $A_i$.

\begin{lemma}
\label{lem:akpw}
$\textsc{AKPW}(G, \delta)$ generates with high probability an AKPW decomposition $\AA$ such that for an edge $e = uv$ with
$l(e) \in [\delta^{i}, \delta^{i + 1})$ and some $j \geq i$,
the probability that $u$ and $v$ are not connected in $A_{j}$ is at most
\begin{align*}
\left( \frac{c_{Partition} \log{n}}{\delta} \right)^{j - i},
\end{align*}
where $c_{Partition}$ is a constant associated with the partition routine.
Furthermore, if $\delta \geq 2 c_{Partition} \log{n}$, it runs
 in expected $\Oh(m \log\log^{1/2}{n})$ time in the RAM model,
 \end{lemma}

\Proof
The termination condition on Line~\ref{line:terminate} implies that $A_s$
is a spanning tree, and the fact that we generate $A_{i + 1}$ by adding
edges to $A_i$ gives $A_i \subseteq A_{i + 1}$.
The bound on diameter can be proven inductively on $i$.

The base case of $i  = 0$ follows from the vertices being singletons,
and as a result having diameter $0$.
For the inductive case, suppose the result is true for $i$.
Then with high probability each connected component in $A_{i + 1}$
corresponds to a tree with diameter $\delta /3$ connecting connected
components in $A_{i}$.
The definition of $E_i$ gives that each of these edges have length
at most $\delta^{i + 1}$, and the inductive hypothesis gives that the
diameter of each connected component in $A_i$ is also at most $\delta^{i + 1}$.
This allows us to bound the diameter of $A_{i + 1}$ by
$(\delta/3) \cdot \delta^{i + 1} + (\delta/3 + 1 )\delta^{i + 1} \leq \delta^{i + 2}$.
Hence the inductive hypothesis holds for $i + 1$ as well.

The guarantees of the probabilistic decomposition routine from
Lemma~\ref{lem:partition} gives that on any level, an edge has its two
endpoints separated with probability $\frac{c_P \log{n}}{\delta}$.
The assumption of the length of $e$ means that it is in $E_i$.
So by the time $A_{j}$ is formed, it has gone through $j - i$ rounds of partition,
and is present iff its endpoints are separated in each of these steps.
Multiplying the probabilities then gives the bound.

If $\delta \geq 2 c_P \log{n}$, then the probability of an edge in $E_i$
appearing in subsequent levels decrease geometrically.
This means that the total expected sizes of $G_t$ processed is $\Oh(m)$.
Combining this with the linear running time of \textsc{Partition} gives
the expected running time once we have the buckets $E_0, E_1,$ etc.
Under the RAM model of computation, these buckets can be formed in
$\Oh(m\log\log^{1/2} {n})$ time using the sorting algorithm by
Han and Thorup~\cite{Han04}.
Incorporating this cost gives the overall running time.
\QED


Combining the bound on diameter and probability of an edge being cut leads
to the bound on the expected $\ell_1$-stretch of an edge
shown by Alon et al.~\cite{AlonKPW95}.
For an edge on the $i$\textsuperscript{th} level, the ratio between its length and the diameter of the 
$j$\textsuperscript{th} level can be bounded by $\delta^{j - i + 1}$.
As $j$ increases, the expected stretch of $e$ then increases by factors of
\begin{align*}
\delta \cdot \Oh \left( \frac{\log{n}}{\delta} \right) = \Oh\left(\log{n}\right),
\end{align*}
which leads to the more than logarithmic bound on the expected
$\ell_1$-stretch.
With $\ell_p$-stretch however, the $p$\textsuperscript{th} power
of the diameter-length ratio only increases by factors of $\delta^{p}$.
This means that, as long as the probabilities of an edge being cut increases
by factors of less than $\delta^{p}$, a better bound can be obtained.


\subsection{AKPW meets Bartal}

In this section, we describe how we combine the AKPW decomposition
and Bartal's scheme into a two-pass algorithm.
At a high level, Bartal's scheme repeatedly partitions the graph in a top-down
fashion, and the choice of having geometrically decreasing diameters translates
to a $\Oh(m\log n)$ running time.
The way our algorithm achieves a speedup is by contracting vertices that are
close to each other, in a way that does not affect the top-down partition scheme.
More specifically, we precompute an appropriate AKPW decomposition, and only
expose a limited number of layers while running the top-down partition.
This way we ensure that each edge only appears in $\Oh(\log\log n)$ calls to
the partition routine.

\begin{figure}[h!t]
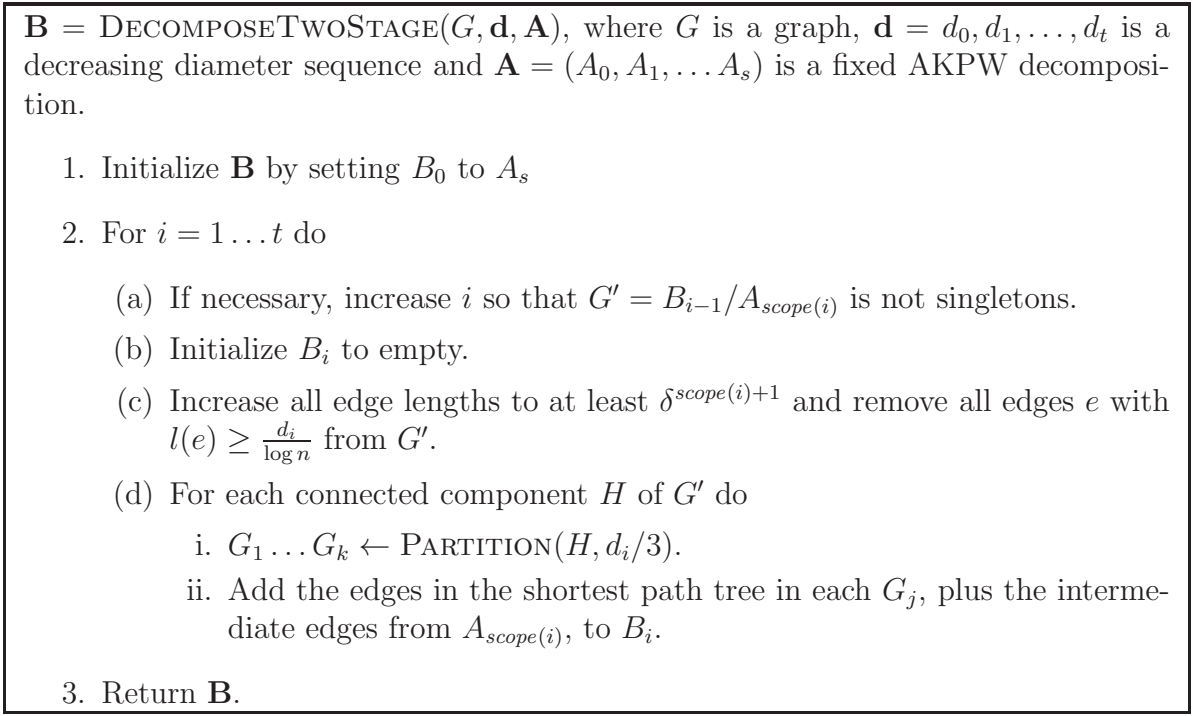

  \vskip 0.2in
  \noindent
  \centering
  \fbox{
    \begin{minipage}{6in}
      \noindent $\BB = \textsc{DecomposeTwoStage}(G, \dd, \AA)$,
      where $G$ is a graph, $\dd = d_0, d_1, \ldots, d_t$ is a decreasing diameter sequence
     and $\AA = (A_0, A_1, \ldots A_s)$ is a fixed AKPW decomposition.
      \begin{enumerate}
	\item Initialize $\BB$ by setting $B_0$ to $A_s$
	\item For $i = 1 \ldots t$ do
		\begin{enumerate}
		        \item If necessary, increase $i$ so that $G'=B_{i-1}/A_{\scope(i)}$ is not singletons.
			\item Initialize $B_i$ to empty.
      \item Increase all edge lengths to at least $\delta^{\scope(i)+1}$
        and remove all edges $e$ with $l(e) \geq \frac{d_{i}}{\log{n}}$ from
        $G'$.
			\item For each connected component $H$ of $G'$ do
				\begin{enumerate}
					\item $G_1 \ldots G_k \leftarrow \textsc{Partition}(H, d_{i} / 3)$.
					\item Add the edges in the shortest path tree in each $G_j$, plus the
            intermediate edges from $A_{\scope(i)}$, to $B_i$.
				\end{enumerate}
		\end{enumerate}
	\item Return $\BB$.
      \end{enumerate}
    \end{minipage}
  }
  \caption{Pseudocode of two pass algorithm for finding
  a Bartal decomposition}
  \label{fig:algo}
\end{figure}

Let $\AA=(A_0,A_1,\cdots,A_s)$ be an AKPW decomposition with parameter
$\delta$, so that $G/A_i$ is the quotient graph where each vertex corresponds to a
cluster of diameter at most $\delta^{i+1}$ in the original graph.
While trying to partition the graph $G$ into pieces of diameter $d$, where
under some notion $d$ is relatively large compared to $\delta^{i+1}$, we
observe that the partition can be done on the quotient graph $G/A_i$ instead.
As the complexity of our partition routine is linear in the number of edges,
there might be some potential gain.
We use the term \emph{scope} to denote the point at which lower levels of the
AKPW decomposition are handled at a coarser granularity.
When the top-down algorithm is reaches diameter $d_i$ in the diameter sequence
$\dd$, this cutoff point in the AKPW decomposition is denoted by
$\scope(i)$.
The algorithm is formalized in Figure~\ref{fig:algo}.

We first show that the increase in edge lengths to
$\delta^{\scope(i) + 1}$ still allows us to bound the
diameter of the connected components of $B_i$.

\begin{lemma}
The diameter of each connected component in $B_i$
is bounded by $d_i$ with high probability.
\end{lemma}

\Proof
By the guarantee of the partition routine, the diameter of each $G_i$ is at
most $\frac{d_i}{3}$ with high probability.
However, since we are measuring diameter of the components in $G$, we also need
to account for the diameter of the components that were shrunken into vertices
when forming $G'$.
These components corresponds to connected pieces in $A_{\scope(i)}$,
therefore the diameters of the corresponding trees are bounded by
$\delta^{\scope(i)+1}$ with high probability.
Our increase of edge lengths in $G'$, on the other hand, ensures that the
length of any edge is more than the diameter of its endpoints.
Hence the total increase in diameter from these pieces is at most twice the
length of a path in $G'$, and the diameter of these components in $G$ can be
bounded by $d_i$.
\QED

Once we established that the diameters of our decomposition is indeed
geometrically decreasing, it remains to bound the probability of an edge
being cut at each level of the decomposition.
In the subsequent sections, we give two different analyses of the algorithm
\textsc{DecomposeTwoStage} with different choices of scope.
We first present a simple version of our algorithm
which ignores a $1/\poly(\log n)$ fraction of the edges, but guarantees
an expected $\ell_1$-stretch close to $\Oh(\log n)$ for rest of the edges.
Then we present a more involved analysis with a careful choice of scope
which leads to a tree with small $\ell_p$-stretch.

\subsection{Decompositions that Ignore $\frac{1}{k}$ of the Edges}
\label{subsec:toss}

In this section, we give a simplified algorithm that ignores $\Oh(\frac1k)$
fraction of the edges, but guarantees for other edges an expected
$\ell_1$-stretch of close to $\Oh(\log n)$.
We also discuss how this relates to the problem of generating low-stretch
subgraphs in parallel and its application to parallel SDD linear system
solvers.

In this simplified algorithm, we use a naive choice of scope, reaching a small power of $k \log n$ into the AKPW decomposition.

Let $\dd=(d_0,d_1,\dots,d_t)$ be a diameter sequence and let
$\AA=(A_0,A_1,\dots,A_s)$ be an AKPW decomposition constructed with parameter $\delta = k\log n$.
We let $\scope(i)=\max\{j \mid \delta^{j+3} \leq d_i \}$.
Note that $\delta^{\scope(i)}$ is always between $\frac{d_i}{\delta^4}$ and $\frac{d_i}{\delta^3}$.
We say an edge $e\in E_i$ is \emph{AKPW-cut} if $e$ is cut in $A_{i+1}$.
Furthermore, we say an edge $e$ is \emph{floating in level $i$} if it exists in $B_{i-1}/A_{\scope(i)}$ and has length less than $\delta^{\scope(i)+1}$.
Note that the floating edges are precisely the edges whose length is increased before running the Bartal decomposition.
We say that an edge is \emph{floating-cut} if it is not AKPW-cut, but is cut by the Bartal decomposition at any level in which it is floating.

The simplification of our analysis over bounding overall $\ell_p$ stretch
is that we can ignore all AKPW-cut or floating-cut edges.
We start by bounding the expected number of edges ignored in
these two ways separately.

\begin{lemma}
  \label{lem:akpwCutEdge}
  Let $\AA=\text{AKPW}(G,\delta)$ where $\delta=k\log n$.
  The expected number of AKPW-cut edges in $\AA$ is at most
  $\Oh(\frac mk)$.
\end{lemma}

\Proof
For an edge $e\in E_i$, the probability that $e$ is cut in $A_{i+1}$ is at
most
\begin{align*}
  \frac{c_{\mit{Partition}}\log n}{\delta}
  =
  \frac{c_{\mit{Partition}}}{k}
\end{align*}
by Lemma~\ref{lem:akpw}, where $c_{\mit{Partition}}$ is the constant associated with the
partition routine.
Linearity of expectation then gives that the expected number of AKPW-cut edges
is at most $\Oh(\frac mk)$.
\QED

We now bound the total number of floating-cut edges:
\begin{lemma}
  \label{lem:floatingEdge}
  The expected number of floating-cut edges is $\Oh(\frac{m}{k})$.
\end{lemma}

\Proof
First, we note that only edges whose length is at least $\frac{d_i}{\delta^4}$ may be floating-cut at level $i$:
any edge smaller than that length that is not AKPW-cut will not be contained in $B_{i-1}/A_{\scope(i)}$.  Furthermore,
by the definition of floating, only edges of lengths at most $\frac{d_i}{\delta^2}$ may be floating.  Therefore,
each edge may only be floating-cut for levels with $d_i$ between $\delta^2$ and $\delta^4$ times the length of the edge.
Since the $d_i$ increase geometrically, there are at most $\log(\delta)$ such levels.

Furthermore, at any given level, the probability that a given edge is floating-cut at the level is at most $\Oh(\frac{\log n}{\delta^2})$,
since any floating edge is passed to the decomposition with length $\frac{d_i}{\delta^2}$.  Taking a union bound over all levels
with $d_i$ between $\delta^2$ and $\delta^4$ times the length of the edge, each edge has at most a $\Oh(\frac{\log n \log \delta}{\delta^2})$
probability of being cut.  Since $\frac{\log \delta}{\delta} = \Oh(1)$, this is $\Oh(\frac{\log n}{\delta}) = \Oh(\frac{1}{k})$.

Again, applying linearity of expectation implies that the expected number of floating-cut edges is $O(\frac{m}{k})$.
\QED

Combining these two bounds gives that the expected number of ignored
edges so far is bounded by $\Oh(\frac{m}{k})$.
We can also check that conditioned on an edge being not ignored,
its probability of being cut on some level is the same as before.

\begin{lemma}
  \label{lem:edgeTossingCutProb}
  Assume $\AA=\text{AKPW}(G,\delta)$.
  We may associate with the output of the algorithm a set of edges $S$,
  with expected size $\Oh(\frac{m}{k})$, such that for any
  edge $e$ with length $l(e)$, conditioned on $e \notin S$,
  is cut on the $i$\textsuperscript{th} level of the Bartal
  decomposition $\BB$ with probability at most
  \begin{align*}
    \Oh\left(\frac{l(e)\log n}{d_i}\right).
  \end{align*}
\end{lemma}

\Proof
We set $S$ to the union of the sets of AKPW-cut and floating-cut edges.

Fix a level $i$ of the Bartal decomposition: if an edge $e$ that is not AKPW-cut
or floating-cut appears in $B_{i-1}/A_{\scope(i)}$, then its length is unchanged.
If $e$ is removed from $G'$ due to $l(e)\ge d_i/\log n$,
the bound becomes trivial.
Otherwise, the guarantees of \textsc{Partition} then give the cut probability.
\QED

\begin{lemma}
  \label{lem:LpIsL1}
  The simplified algorithm produces with high
probability an embeddable Bartal decomposition with diameters
bounded by $\dd$ where all but (in expectation) $\Oh(\frac mk)$ edges satisfy
$\expct_\BB[\str_{\BB, \dd}(e)] \leq \Oh(\log n(\log(k \log n))^2)$.
\end{lemma}

\Proof
Let $p=1-1/\log(k \log n)$ and $q=(1+p)/2$.
Applying Lemma~\ref{lem:edgeTossingCutProb} and Lemma~\ref{lem:goodProb}
we get that for edges not in $S$,
$\expct_\BB[\str_{\BB,\dd}^q(e)]=\Oh(\log^qn\log(k \log n))$.
Then using \textsc{EmbeddableDecompose} as a black box we obtain an embeddable
decomposition with expected $l_p$-stretches of $\Oh(\log^pn (\log(k\log n))^2)$ for
non-removed edges.

By repeatedly running this algorithm, in an expected constant number
of iterations, we obtain an embeddable decomposition $\BB$ with
diameters bounded by $\dd$ such that for a set of edges
$E' \subseteq E$ and $|E'| \geq m - \Oh(\frac{m}{k})$ we have:
\[
\sum_{e \in E'} \expct_\BB[\str_{\BB,\dd}^q(e)]=\Oh(m \log^q n(\log(k\log n))^2).
\]
By Markov's inequality, at most $1/k$ of the edges in $E'$
can have $\str_{\BB,\dd}^q (e) \geq \Oh(k \log^q n(\log(k\log n))^2)$.
This gives a set of edges $E''$ with size at least
$m - \Oh(\frac{m}{k})$ such that any edge $e \in E''$ satisfies
$\str_{\BB, \dd}^q(e) \leq \Oh(k \log^q n(\log(k\log n))^2) \leq \Oh((k \log n)^2)$.

But for each of these edges
\begin{align*}
\str_{\BB, \dd}(e) &= (\str_{\BB, \dd}^q(e))^{1/q} \\
& \leq (\str_{\BB, \dd}^q(e))^{1+2/\log(k \log n)} \\
& \leq  \str_{\BB, \dd}^q(e) \cdot
	\Oh \left( \left( k \log n\right)^{4 /\log(k \log n)} \right) \\
&= \Oh(\str_{\BB, \dd}^q(e)).
\end{align*}

Excluding these high-stretch edges, the $\ell_1$ stretch is thus
at most a constant factor worse than the $\ell_q$ stretch,
and can be bounded by $\Oh(\log n(\log(k \log n))^2)$.
\QED

The total running time of \textsc{DecomposeTwoStage}
is dominated by the calls to \textsc{Partition}.
The total cost of these calls can be bounded by the
expected number of calls that an edge participates in.

\begin{lemma}
  \label{lem:edgeTossTime}
  For any edge $e$, the expected number of iterations in which $e$ appears
  is bounded by $\Oh(\log(k\log n))$.
\end{lemma}

\Proof
As pointed out in the proof of \ref{lem:floatingEdge},
an edge that is not AKPW-cut only appears in level $i$ of the Bartal decomposition
if $l(e)\in[\frac{d_i}{\delta^5},\frac{d_i}{\log n})$.
Since the diameters decrease geometrically, there are at most $\Oh(\log(k \log n))$
such levels.
AKPW-cut edges can appear sooner than other edges from the
same weight bucket, but using an argument similar to the proof of
Lemma~\ref{lem:akpwCutEdge} we observe that the edge propagates up $j$ levels
in the AKPW decomposition with probability at most $(\frac1k)^j$.
Therefore the expected number of such appearances by an APKW-cut edge is at most
$\sum_i(\frac1k)^i=\Oh(1)$.
\QED

Combining all of the above we obtain the following result about our simplified
algorithm.
The complete analysis of its running time is deferred to
Section~\ref{subsec:iCanHazTree}.

\begin{lemma}
  \label{lem:edgeTossLemma}
For any  $k$, given an AKPW decomposition $\AA$ with $\delta = k \log{n}$,
we can find in $\Oh(m\log(k \log n))$ time
an embeddable Bartal decomposition
such that all but expected $\Oh(\frac mk)$ edges
have expected total $\ell_1$-stretch of at most $\Oh(m \log n(\log(k \log n))^2)$.
\end{lemma}

\subsubsection{Parallelization}
\label{subsubsec:parallel}

If we relax the requirement of asking for a tree, the above analysis shows that
we can obtain low stretch subgraphs edges and total stretch
of $\Oh(\log n(\log(k \log n))^2)$ for all but $\Oh(\frac{m}{k})$ edges.
As our algorithmic primitive \textsc{Partition} admits parallelization~\cite{MillerPX13},
we also obtain a parallel algorithm for constructing low stretch subgraphs.
These subgraphs are used in the parallel SDD linear system solver
by~\cite{BlellochGKMPT11}.
By observing that \textsc{Partition} is run on graphs with edge weights
within $\delta$ of each other and hop diameter at most polynomial in $\delta=k\log n$,
and invoking tree-contraction routines to extract the final
tree~\cite{MillerR89:book}, we can obtain the following result.

\begin{lemma}
  \label{lem:parallel}
For any graph $G$ with polynomially bounded edge weights
and $k \leq \poly(\log n)$, in $\Oh(k\log^2n\log\log n)$ depth and
$\Oh(m\log n)$ work we can generate an embeddable tree
of size $\Oh(n)$ such that the total $\ell_1$-stretch of all but
$\Oh(\frac{m}{k})$ edges of $G$ is $\Oh(m \log n(\log(k\log n))^2)$.
\end{lemma}

\subsection{Bounding Expected $\ell_p$-Stretch of Any Edge}

In this section we present our full algorithm and bound the
expected $\ell_p$-stretch of all edges
Since we can no longer ignore edges whose lengths we increase
while performing the top-down partition, we need to
choose the scope carefully in order to control their probability
of being cut during the second stage of the algorithm.
We start off by choosing a different $\delta$ when computing
the AKPW decomposition.

\begin{lemma}
\label{lem:akpwCutProb}
If $\AA$ is generated by a call to $\textsc{AKPW}(G, \delta)$
with $\delta \geq \left(c_P \log{n} \right)^{\frac{1}{1 - q}}$,
then the probability of an edge $e \in E_i$ being cut in level $j$
is at most $\delta^{-q (j - i)}$.
\end{lemma}

\Proof
Manipulating the condition gives $c_P \log{n} \leq \delta^{1 - q}$, and
therefore using Lemma~\ref{lem:akpw} we can bound the probability by
\begin{align*}
\left( \frac{c_P \log{n}}{\delta} \right)^{j - i}
& \leq \left( \frac{\delta^{1 - q}}{\delta} \right)^{j - i}
= \delta^{-q (j - i)}.
\end{align*}
\QED

Since $\delta$ is $\poly(\log{n})$, we can use this bound to show that expected
$\ell_p$-stretch of an edge in an AKPW-decomposition can be bounded by
$\poly(\log{n})$.
The exponent here can be optimized by taking into account
the trade-offs given in Lemma~\ref{lem:goodProb}.

This extra factor of $\delta$ can also be absorbed into the
analysis of Bartal decompositions.
When $l(e)$ is significantly less than $d$,
the difference between $\frac{l(e) \log{n}}{d}$ and
$\left(\frac{l(e) \log{n}}{d}\right)^{q}$ is more than $\delta$.
This means that for an floating edge that originated much lower in
the bucket of the AKPW decomposition,
we can afford to increase its probability of being cut by a factor of $\delta$.

From the perspective of the low-diameter decomposition routine,
this step corresponds to increasing the length of an edge.
This increase in length can then be used to bound the diameter
of a cluster in the Bartal decomposition, and also ensures that all edges that
we consider have lengths close to the diameter that we partition into.
On the other hand, in order to control this increase in lengths, and in turn to
control the increase in the cut probabilities, we need to use a different scope
when performing the top-down decomposition.

\begin{definition}
\label{def:scope}
For an exponent $q$ and a parameter $\delta \geq \log{n}$,
we let the scope of a diameter $d$ be
\begin{align*}
  \scope(i) := \max_i \left\{ \delta^{i + \frac{1}{1 - q} + 1} \leq d_i\right\}.
\end{align*}
\end{definition}

Note that for small $d$, $\scope(i)$ may be negative.
As we will refer to $A_{\scope(i)}$, we assume that $A_i = \emptyset$ for $i < 0$.
Our full algorithm can then be viewed as only processing the 
edges within the scope using Bartal's top-down algorithm.
Its pseudocode is given in Figure~\ref{fig:algo}.

Note that it is not necessary to perform explicit contraction and expansion of the AKPW clusters in every recursive call.
In an effective implementation, they can be expanded gradually, as $\scope(i)$ is monotonic in $d_i$.

The increase in edge lengths leads to
increases in the probabilities of edges being cut.
We next show that because the AKPW decomposition is computed 
using a higher norm, this increase can be absorbed, giving
a probability that is still closely related to the
$p$\textsuperscript{th} power of the ratio between
the current diameter and the length of the edge.

\begin{lemma}
\label{lem:fullCutProb}
Assume $\AA = \textsc{AKPW}(G, \delta)$ with parameter specified as above.
For any edge $e$ with length $l(e)$ and any level $i$,
the probability that $e$ is cut at level $i$ of
$\BB=\textsc{DecomposeTwoStage}(G,\dd,\AA)$ is
    \begin{align*}
        \Oh\left(\left(\frac{l(e) \log n}{d_i}\right)^{q}\right).
    \end{align*}
\end{lemma}

\Proof
There are two cases to consider based whether the length of the edge is more
than $\delta^{\scope(i) + 1}$.
If it is and it appears in $G'$, then its length is retained.
The guarantees of $\textsc{Partition}$ then gives that it is cut with probability
\begin{align*}
  \Oh\left(\frac{l(e) \log n}{d_i}\right)
\leq \Oh\left(\left(\frac{l(e) \log n}{d_i}\right)^q\right),
\end{align*}
where the inequality follows from $l(e) \log{n} \leq d_i$.

Otherwise, since we contracted the connected components in
$A_{\scope(i)}$, the edge is only cut at level $i$ if
it is both cut in $A_{\scope(i)}$ and cut by the partition routine.
Lemma~\ref{lem:akpwCutProb} gives that if the edge is from $E_j$, its
probability of being cut in $A_{\scope(i)}$
can be bounded by $\delta^{-q(\scope(i) - j)}$.
Combining with the fact that $\delta^{j} \leq l(e)$ allows us to bound this probability by
\begin{align*}
  \left( \frac{l(e)}{\delta^{\scope(i)}}\right)^{q}.
\end{align*}
Also, since the weight of the edge is set to $\delta^{\scope(i) + 1}$ in $G'$,
its probability of being cut by $\textsc{Partition}$ is
\begin{align*}
  \Oh\left(\frac{\delta^{\scope(i) + 1}  \log n}{d_i}\right).
\end{align*}
As the partition routine is independent of the AKPW decomposition routine,
the overall probability can be bounded by
\begin{align*}
  \Oh\left(\frac{\delta^{\scope(i) + 1}  \log n}{d_i} \cdot
  \left( \frac{l(e)}{\delta^{\scope(i)}}\right)^{q}
\right)
& = \Oh\left(
\left( \frac{l(e) \log{n}}{d_i}\right)^{q}
\cdot \delta \log^{1 - q}{n} \cdot
\left( \frac{\delta^{scope(i)}}{d_i} \right)^{1 - q} \right).
\end{align*}
Recall from Definition~\ref{def:scope} that $scope(i)$
is chosen to satisfy $\delta^{scope(i) + \frac{1}{1 - q} + 1} \leq d_i$.
This along with the assumption that $\delta \geq \log{n}$ gives
\begin{align*}
\delta \log^{1 - q}{n} \cdot
\left( \frac{\delta^{scope(i)}}{d_i} \right)^{1 - q}
\leq \delta^{2 - q} \left( \delta^{-\frac{2 - q}{1 - q}} \right)^{1 - q}
\leq 1.
\end{align*}
Therefore, in this case the probability of $e$ being
cut can also be bounded by $\Oh\left(\left(\frac{l(e) \log n}{d_i}\right)^q\right)$.
\QED

Combining this bound with Lemma~\ref{lem:goodProb} and
setting $q = \frac{1 + p}{2}$ gives the bound on $\ell_p$-stretch.

\begin{corollary}
\label{cor:twoStageStr}
If $q$ is set to $\frac{1 + p}{2}$, we have for any edge $e$
$\expct_{\BB}[\str_{\BB, \dd}(e)] \leq \Oh( \frac{1}{1 - p} \log^{p}n)$.
\end{corollary}

Therefore, we can still obtain the properties of a good Bartal decomposition
by only considering edges in the scope during the top-down partition process.
On the other hand, this shrinking drastically improves the performance of our algorithm.

\begin{lemma}
\label{lem:edgeLevelCount}
Assume $\AA = \textsc{AKPW}(G, \delta)$.
For any edge $e$, the expected number of iterations of $\textsc{DecomposeTwoStage}$ in which $e$ is included in the
graph given to $\textsc{Partition}$ can be bounded by $\Oh(\frac{1}{1 - p} \log \log n).$
\end{lemma}
\Proof
Note that for any level $i$ it holds that
\begin{align*}
    \delta^{scope(i)} \geq d_i \delta^{-\frac{1}{1-q}-2}.
\end{align*}
Since the diameters of the levels decrease geometrically,
there are at most $\Oh(\frac{1}{1 - q} \log \log n)$ levels $i$ such that $l(e) \in [d_i\delta^{-\frac{1}{1-q}-2}, \frac{d_i}{\log n})$.

The expected number of occurrences of $e$ in lower levels
can be bounded using Lemma~\ref{lem:akpwCutProb}
in a way similar to the proof of the above Lemma.
Summing over all the levels $i$ where $e$ is in a lower level gives:
\[
    \sum_{i: l(e) < d_i\delta^{-\frac{1}{1-q}-2}} \left(\frac{l(e)}{\delta^{scope(i)}}\right)^q
\]

Substituting in the bound on $\delta^{scope(i)}$ from above and
rearranging then gives:
\[
 \leq \sum_{i: l(e) \leq d_i\delta^{-\frac{1}{1-q}-2}} \left(\frac{l(e)}{d_i}\delta^{\frac{1}{1-q}+2}\right)^q.
\]
As $d_i$ increase geometrically, this is a geometric sum
with the first term at most $1$.
Therefore the expected number of times that $e$ appears on
some level $i$ while being out of scope is $\Oh(1)$.
\QED

Recall that each call to $\textsc{Partition}$ runs in time linear in the
number of edges.
This then implies  a total cost of $\Oh(m\log\log{n})$ for all the partition steps.
We can now proceed to extract a tree from this decomposition, and analyze
the overall running time.

\subsection{Returning a Tree}
\label{subsec:iCanHazTree}

We now give the overall algorithm and analyze its performance.
Introducing the notion of scope in the recursive algorithm limits each edge to
appear in at most $\Oh(\log\log{n})$ levels.
Each of these calls partitions $G'$ in time linear in its size,
which should give a total of $\Oh(m \log\log{n})$.
However, the goal of the algorithm as stated is to produce a Bartal decomposition, which has a spanning tree at each level.
Explicitly generating this gives a total size of $\Omega(n t)$, where $t$ is the number of recursive calls.
As a result, we will circumvent this by storing only an implicit representation
of the Bartal decomposition to find the final tree.

This smaller implicit representation stems from the observation that
large parts of the $B_i$s are trees from the AKPW decomposition, $A_i$.
As a result, such succinct representations are possible
if we have pointers to the connected components of $A_i$.
We first analyze the quality and size of this implicit decomposition,
and the running time for producing it.

\begin{figure}[ht]
\vskip 0.2in
\centering
\fbox{
\begin{minipage}{6in}
\noindent $\BB, \dd = \textsc{Decompose} (G, p)$, where $G$ is a graph, $p$ is an exponent
\begin{enumerate}
	\item Set $q = \frac{1 + p}{2}$, $\delta = \left(c \log{n} \right)^{\frac{1}{q - p}}$.
	\item Compute an AKPW decomposition of $G$, $\AA = \textsc{AKPW}(G, \delta)$.
	\item Let $\dd=(d_0,d_1,\cdots,d_t)$ be a geometrically decreasing sequence
    diameters where $d_0$ is the diameter of $A_s$.
	\item Set $\BB := \textsc{DecomposeTwoStage}(G, \dd,\AA)$.
	\item Set $B_0$ to $A_s$.
	\item Return $\BB, \dd$.
\end{enumerate}
\end{minipage}
}
\caption{Overall decomposition algorithm}
\label{fig:driver}
\end{figure}

\begin{lemma}
\label{lem:implicitDecomp}
There is a routine that for any graph $G$ and and parameter $p < 1$,
produces in expected $\Oh(\frac{1}{1 - p} m\log\log{n})$ time
an implicit representation of a Bartal decomposition $\BB$
with expected size $\Oh(\frac{1}{1 - p} m\log\log{n})$
and diameter bounds $\dd$ such that with high probability:
\begin{itemize}
	\item $\BB$ is embeddable into $G$, and
	\item for any edge $e$, $\expct_{\BB}(\str_{\BB, \dd}^{p}(e)) \leq
    \Oh((\frac{1}{1 - p})^2 \log^{p}n)$.
	\item $\BB$ consist of edges and weighted connected
		components of an AKPW decomposition
\end{itemize}
\end{lemma}

\Proof
Consider calling \textsc{EmbeddableDecompose}  from Section~\ref{subsec:embed}
with the routine given in Figure~\ref{fig:driver}.
The properties of $\BB$ and the bounds on stretch follows from
Lemma~\ref{lem:stretchpq} and Corollary~\ref{cor:twoStageStr}.

Since the number of AKPW components implicitly referred to at each level
of the recursive call is bounded by the total number of vertices,
and in turn the number of edges,
the total number of such references is bounded by the size of the $G'$s as well.
This gives the bound on the size of the implicit representation.

We now bound the runnign time.
In the RAM model, bucketing the edges and computing
the AKPW decomposition can be done in $\Oh(m \log{\log{n}})$ time.
The resulting tree can be viewed as a laminar decomposition of the graph.
This is crucial for making the adjustment in \textsc{DecomposeTwoStage} in
$\Oh(1)$ time to ensure that $A_{scope(i)}$ is disconnected.
As we set $q$ to $\frac{1 + p}{2}$, by Lemma~\ref{lem:edgeLevelCount},
each edge is expected to participate in $\Oh(\frac{1}{1 - p} \log\log{n})$ recursive calls,
which gives a bound on the expected total.

The transformation of the edge weights consists of a linear-time
pre-processing, and scaling each level by a fixed parameter in
the post-post processing step.
This process affects the implicit decomposition by changing the weights
of the AKPW pieces, which is can be done implicitly in $\Oh(1)$ time by
attaching extra `flags' to the clusters.
\QED


It remains to show that an embeddable tree can be generated efficiently from this
implicit representation.
To do this, we define the notion of a contracted tree with respect to a subset
of vertices, obtained by repeating the two combinatorial steps that preserve
embeddability described in Section~\ref{sec:background}.
\begin{definition}
\label{def:contraction}
We define the \emph{contraction} of a tree $T$ to a subset of its vertices $S$
as the unique tree arising from repeating the following operations while
possible:
\begin{itemize}
    \item removal of a degree 1 vertex not in $S$, and
    \item contraction of a degree 2 vertex not in $S$.
\end{itemize}
\end{definition}
We note that it is enough to find contractions of the trees from the AKPW
decomposition to the corresponding sets of connecting endpoints in the implicit
representation.
Here we use the fact that the AKPW decomposition is in fact a single tree.

\begin{fact}
\label{fact:akpw_contr}
    Let $\AA = A_0, \ldots, A_s$ be an AKPW decomposition of $G$.
    Let $S$ be a subset of vertices of $G$.
    For any $i$ in $\{0, \ldots, s\}$, if $S$ is contained in a single connected component of $A_i$, then
    the contraction of $A_i$ to $S$ is equal to the contraction of $A_s$ to $S$.
\end{fact}

This allows us to use data structures to find the contractions of the AKPW trees to the respective vertex sets
more efficiently.

\begin{lemma}
\label{lem:lca}
    Given a tree $A_s$ on the vertex set $V$ (with $|V| = n$) and subsets $S_1, \ldots, S_k$ of $V$ of total size $\Oh(n)$, we can
    generate the contractions of $A_s$ to each of the sets $S_i$ in time
    $\Oh(n)$ in the RAM model and $\Oh(n \alpha(n))$ in the pointer machine model.
\end{lemma}
\Proof
    Root $A_s$ arbitrarily.
    Note that the only explicit vertices required in the contraction of $A_s$ to a set $S \subseteq V$ are
    \begin{align*}
        \Gamma(S) \defeq S \cup \{LCA(u, v): u, v \in S\}
    \end{align*}
    where $LCA(u, v)$ denotes the lowest common ancestor of $u$ and $v$ in $A_s$.
    Moreover, it is easily verified that if we sort the vertices $v_1, \ldots, v_{|S|}$ of $S$ according
    to the depth first search pre-ordering, then
    \begin{align*}
        \Gamma(S) = S \cup \{LCA(v_i, v_{i + 1}) : 1 \leq i < |S|\}.
    \end{align*}
    We can therefore find $\Gamma(S_i)$ for each $i$ simultaneously in the following steps:
    
    \begin{enumerate}
        \item Sort the elements of each $S_i$ according to the pre-ordering, using a single depth-first search traversal of $A_s$.
        \item Prepare a list of lowest common ancestor queries for each pair of vertices adjacent in the sorted order in each set $S_i$.
        \item Answer all the queries simultaneously using an off-line lowest common ancestor finding algorithm.
    \end{enumerate}
    Since the total number of queries in the last step is $\Oh(n)$, its running time is $\Oh(n \alpha(n))$
    in the pointer machine model using disjoint union~\cite{Tarjan79}, and $\Oh(n)$
    in the RAM model~\cite{GabowTarjan83}.

    Once we find the sets $\Gamma(S_i)$ for each $i$, we can reconstruct the contractions of $A_s$ as follows:
    \begin{enumerate}
        \item Find the full traversal of the vertices in $\Gamma(S_i)$ for each $i$, using a single depth first search traversal of $A_s$.
        \item Use this information to reconstruct the trees~\cite{Vuillemin80}.
    \end{enumerate}
\QED

Applying this procedure to the implicit decomposition then
leads to the final embeddable tree.

\Proofof{Theorem~\ref{thm:main}}
 Consider the distribution over Bartal decompositions given by
Lemma~\ref{lem:implicitDecomp}.
We will apply the construction given in Lemma~\ref{lem:buildTree},
albeit in a highly efficient manner.

For the parts of the decomposition that are explicitly given, the routine
runs in linear time.
The more intricate part is to extract the smaller contractions
from the AKPW components that are referenced to implicitly.
Since all levels of the AKPW decomposition are subtrees of $A_s$,
these are equivalent to finding contractions of $A_s$
for several sets of vertices, as stated in Fact~$\ref{fact:akpw_contr}$.
The algorithm given in Lemma~\ref{lem:lca} performs this operation in linear time.
Concatenating these trees with the one generated from the explicit part of
the decomposition gives the final result.
\QED


\section*{Acknowledgments}

We thank Ittai Abraham, Anupam Gupta, Nick Harvey, Jon Kelner, Yiannis Koutis, Aaron Sidford, Dan Spielman, and Zeyuan Zhu for very helpful comments and discussions.

\newcommand{\etalchar}[1]{$^{#1}$}

    \bibliographystyle{alpha}

\begin{appendix}

\section{Sufficiency of Embeddability}
\label{sec:embedOk}

In the construction of our trees, we made a crucial relaxation of only requiring embeddability, rather than restricting to subgraphs.
In this section, we show that linear operators on the resulting graph
can be related to linear operators on the original graph.
Our analysis is applicable to $\ell_\infty$ flows as well.
%

The spectral approximation of two graphs can be defined in terms
of their Laplacians.
As we will interpret these objects combinatorially, we omit their
definition and refer the reader to Doyle and Snell~\cite{DoyleS84:book}.
For matrices, we can define a partial ordering $\preceq$ where
$\mata \preceq \matb$ if $\matb - \mata$ is positive semidefinite.
That is, for any vector $\vecx$ we have
\[
\vecx^T \mata \vecx \leq \vecx^T \matb \vecb.
\]

If we let the graph formed by adding the tree to $G$ be $H$,
then our goal is to bound $\laplacian_G$ and $\laplacian_H$
with each other.
Instead of doing this directly, it is easier to relate their pseudoinverses.
This will be done by interpreting $\vecx^T \laplacian^{\dag} \vecx$
in terms of the energy of electrical flows.
The energy of an electrical flow is defined as the
sum of squares of the flows on the edges multiplied by their resistances,
which in our case are equal to the lengths of the edges.
Given a flow $f \in \Re^{E}$, we will denote its electrical energy using
\begin{align*}
	\mathcal{E}_{G}(f) \defeq \sum_{e} l_e f(e)^2.
\end{align*}

The residue of a flow $f$ is the net in/out flow at each vertex.
This give a vector on all vertices, and finding the minimum energy of flows
that meet a given residue is equivalent to computing $\vecx^T \laplacian^{\dag} \vecx$.
The following fact plays a central role in the monograph by
Doyle and Snell~\cite{DoyleS84:book}:
\begin{fact}
\label{fact:energy}
Let $G$ be a connected graph.
For any vector $\vecx$ orthogonal to the all ones vector,
$\vecx^T \laplacian_G^{\dag} \vecx$ equals the minimum
electrical energy of a flow with residue $\vecx$.
\end{fact}

\begin{lemma}
\label{lem:steinerOk}
Let $G = (V_G, E_G, \weight_G)$ and $H = (V_H, E_H, \weight_H)$ be graphs such that $G$ is a subgraph of $H$
in the weighted sense and $H \setminus G$ is embeddable in $G$.
Furthermore, let the graph Laplacians of $G$ and $H$
be $\laplacian_G$ and $\laplacian_H$ respectively.
Also, let $\proj$ be the $|V_G| \times |V_H|$ matrix
with one $1$ in each row at the position that vertex
corresponds to in $H$ and $0$ everywhere else, and $\proj_1$
the orthogonal projection operator onto the part of
$\Re^{V_G}$ that's orthogonal to the all-ones vector.
Then we have:
\begin{align*}
\frac{1}{2} \laplacian_G^{\dag}
\preceq \proj_1 \proj \laplacian_H^{\dag} \proj^T \proj_1^T \preceq
\laplacian_G^{\dag}.
\end{align*}
\end{lemma}

\Proof
Since $\proj_1^T = \proj_1$ projects out any part space spanned by the all ones vector,
and is this precisely the null space of $\laplacian_G$, it suffices to show
the result for all vectors $\vecx_G$ orthogonal to the all-1s vector.
These vectors are in turn valid demand vectors for electrical flows.
Therefore, the statement is equivalent to relating the minimum
energies of electrical flows routing $\vecx_G$ on $G$ and
$\proj^T \vecx_G$ on $H$.

We first show that flows on $H$ take less energy than the ones in $G$.
Let $\vecx_G$ be any vector orthogonal to the all ones vector,
and $f_G^{*}$ be the flow of minimum energy in $G$ that meets demand $\vecx_G$.
Setting the same flow on the edges of $E(G)$ in $H$ and $0$
on all other edges yields a flow $f_H$.
The residue of this flow is the same residue in $V_G$,
and $0$ everywhere else, and therefore equal to $\proj^T \vecx_G$.
Since $G$ is a subgraph of $H$ in the weighted sense,
the lengths of these edges can only be less.
Therefore the energy of $f_H$ is at most the energy of $f_G$ and we have
\[
\vecx_G^T \proj \laplacian_H^{\dag} \proj^T \vecx_G
\leq \mathcal{E}_{H}(f_H)
\leq \mathcal{E}_{G}(f_G^{*})
= \vecx_G^T \laplacian_G^{\dag} \vecx_G.
\]

For the reverse direction, we use the embedding of $H \setminus G$ into $G$
to transfer the flow from $H$ into $G$.
Let $\vecx_G$ be any vector orthogonal to the all ones vector,
and $f_H^{*}$ the flow of minimum energy in $H$ that has residue $\proj^T \vecx_G$.
This flow can be transformed into one in $G$ that has residue $\vecx_G$ using the embedding.
Let vertex/edge mapping of this embedding be $\pi_V$ and $\pi_E$ respectively.

If an edge $e \in E_H$ is also in $E_G$, we keep its flow value in $G$.
Otherwise, we route its flow along the path that the edge is mapped to.
Formally, if the edge is from $u$ to $v$, $f_H(e)$ units of flow is routed
from $\pi_V(u)$ to $\pi_V(v)$ along $path(e)$.
We first check that the resulting flow, $f_G$ has residue $\vecx_G$.
The net amount of flow into a vertex $u \in V_G$ is
\[
\sum_{uv \in E_G} f_H^{*}(e)
+ \sum_{u'v' \in E_H \setminus E_G, \pi_V(u') = u} f_H^{*}(e)\\
= \sum_{uv \in E_G} f_H^{*}(e) + \sum_{u' \in V_H, \pi_V(u') = u} \left( \sum_{u'v' \in E_H \setminus E_G} f_H^{*}(e) \right).\\
\]
Reordering the summations and noting that $\Pi(u) = u$ gives
\[
= \sum_{u' \in V_H, \pi_V(u') = u} \sum_{u'v' \in E_H} f_H(e)
= \sum_{u' \in V_H, \pi_V(u') = u} \left( \proj^T \vecx_G \right)(e)
= x_G(u).
\]
The last equality is because $\pi_V(u) = u$,
and all vertices not in $V_G$ having residue $0$ in $\proj^T \vecx_G$.

To bound the energy of this flow, the property of the embedding gives that
if split the edges of $G$ into the paths that form the embedding,
each edge is used at most once.
Therefore, if we double the weights of $G$, we can use one copy to support $G$,
and one copy to support the embedding.
The energy of this flow is then the same.
Hence there is an electrical flow $f_G$ in $G$ such that $\mathcal{E}_G(f_G) \leq 2 \mathcal{E}_H(f_H^{*})$.
Fact~\ref{fact:energy} then gives that it is an upper bound for
$\vecx_G^T \laplacian_G^{\dag} \vecx_G$, completing the proof.
\QED

\end{appendix}

\end{document}